\documentclass[reprint,
 superscriptaddress,
 amsmath,
 amssymb,
 nofootinbib,
 aps,
 pra,
]{revtex4-2}
\usepackage[utf8]{inputenc}
\usepackage{graphicx}% Include figure files
\usepackage{dcolumn}% Align table columns on decimal point
\usepackage{bm}% bold math
\usepackage{natbib}
\usepackage{dirtytalk}
\usepackage{csquotes}
\usepackage{textgreek}
\usepackage[caption=false, position=bottom, justification = justified, singlelinecheck = false]{subfig}
\usepackage{adjustbox}
\usepackage{placeins}
\usepackage{csquotes}
\usepackage{amsfonts}
\usepackage{booktabs}
\begin{document}

%\preprint{APS/123-QED}
\title{Element-specific, non-destructive profiling of layered heterostructures}% Force line breaks with \\

\author{Nicolò D'Anna}
\email{ndanna@ucsd.edu}
\altaffiliation{Present address: Department of Physics
University of California San Diego, La Jolla, CA 92093, USA}
\affiliation{PSI Center for Photon Science, Paul Scherrer Institute, 5232 Villigen PSI, Switzerland}
\affiliation{Laboratory for Solid State Physics and Quantum Center, ETH Zurich, Zurich, Switzerland}
\author{Jamie Bragg}
\affiliation{London Centre for Nanotechnology, University College London, WC1H 0AH, London, UK}
\affiliation{Department of Electronic and Electrical Engineering, University College London, London WC1E 7SE, UK}
\author{Elizabeth Skoropata}
\author{Nazareth Ortiz Hernández}
\affiliation{PSI Center for Photon Science, Paul Scherrer Institute, 5232 Villigen PSI, Switzerland}
\author{Aidan~G.~McConnell}
\author{Maël~Clémence}
\affiliation{PSI Center for Photon Science, Paul Scherrer Institute, 5232 Villigen PSI, Switzerland}
\affiliation{Laboratory for Solid State Physics and Quantum Center, ETH Zurich, Zurich, Switzerland}
\author{Hiroki~Ueda}
\affiliation{PSI Center for Photon Science, Paul Scherrer Institute, 5232 Villigen PSI, Switzerland}
\author{Procopios~C.~Constantinou}
\affiliation{PSI Center for Photon Science, Paul Scherrer Institute, 5232 Villigen PSI, Switzerland}
\affiliation{London Centre for Nanotechnology, University College London, WC1H 0AH, London, UK}
\affiliation{Department of Physics and Astronomy, University College London, WC1E 6BT, London, UK}
\author{Kieran Spruce}
\affiliation{London Centre for Nanotechnology, University College London, WC1H 0AH, London, UK}
\affiliation{Department of Electronic and Electrical Engineering, University College London, London WC1E 7SE, UK}
\author{Taylor~J.Z.~Stock}
\affiliation{London Centre for Nanotechnology, University College London, WC1H 0AH, London, UK}
\affiliation{Department of Electronic and Electrical Engineering, University College London, London WC1E 7SE, UK}
\author{Sarah Fearn}
\affiliation{London Centre for Nanotechnology, University College London, WC1H 0AH, London, UK}
\affiliation{Department of Materials, Imperial College of London, London SW7 2AZ, UK}
\author{Steven~R.~Schofield}
\affiliation{London Centre for Nanotechnology, University College London, WC1H 0AH, London, UK}
\affiliation{Department of Physics and Astronomy, University College London, WC1E 6BT, London, UK}
\author{Neil~J.~Curson}
\affiliation{London Centre for Nanotechnology, University College London, WC1H 0AH, London, UK}
\affiliation{Department of Electronic and Electrical Engineering, University College London, London WC1E 7SE, UK}
\author{Dario~Ferreira~Sanchez}
\author{Daniel Grolimund}
\author{Urs Staub}
\author{Guy Matmon}
\author{Simon Gerber}
\affiliation{PSI Center for Photon Science, Paul Scherrer Institute, 5232 Villigen PSI, Switzerland}
\author{Gabriel Aeppli}
\email{gabriel.aeppli@psi.ch}
\affiliation{PSI Center for Photon Science, Paul Scherrer Institute, 5232 Villigen PSI, Switzerland}
\affiliation{Laboratory for Solid State Physics and Quantum Center, ETH Zurich, Zurich, Switzerland}
\affiliation{Institute of Physics, EPF Lausanne, 1015 Lausanne, Switzerland}

\begin{abstract}
 \textbf{Fabrication of semiconductor heterostructures is now so precise that metrology has become a key challenge for progress in science and applications. It is now relatively straightforward to characterize classic III-V and group IV heterostructures consisting of slabs of different semiconductor alloys with thicknesses of $\sim$5~nm and greater using sophisticated tools such as \mbox{X-ray} diffraction, high energy X-ray photoemission spectroscopy, and secondary ion mass spectrometry. However, profiling thin layers with nm or sub-nm thickness, \textit{e.g.} atomically thin dopant layers (\mbox{$\delta$-layers}), of impurities required for modulation doping and spin-based quantum and classical information technologies is more challenging. 
 Here, we present theory and experiment showing how resonant-contrast \mbox{X-ray} reflectometry meets this challenge. The technique takes advantage of the change in the scattering factor of atoms as their core level resonances are scanned by varying the \mbox{X-ray} energy. 
We demonstrate the capability of the resulting element-selective, non-destructive profilometry for single arsenic $\delta$-layers within silicon, and show that the sub-nm electronic thickness of the \mbox{$\delta$-layers} corresponds to sub-nm chemical thickness. In combination with X-ray fluorescence imaging, this enables non-destructive three-dimensional characterization of nano-structured quantum devices.
Due to the strong resonances at soft X-ray wavelengths, the technique is also ideally suited to characterize layered quantum materials, such as cuprates or the topical infinite-layer nickelates.}
\end{abstract}

\maketitle

%intro
With the advent of scanning tunneling microscopy~(STM) lithography in the 1990s \cite{STM_manip,STM_corrals}, it became possible to fabricate dopant-based nano-electronic structures in semiconductors \cite{STM_P}, as can now be done with phosphorus, arsenic, and boron in silicon \cite{Single_atom_transistor,Stock_As_count,Dwyer:2021aa}. In the meantime, industrially fabricated transistors have reached a 7~nm scale \cite{Samsung_FET}.
A number of methods to image electronic nano-structures are available. The most popular are destructive, and include transmission electron microscopy~\cite{Wang2018}, atom-probe tomography \cite{atome_probe, Wang2018} and secondary ion mass spectrometry (SIMS)~\cite{SIMS, Wang2018}.
Non-destructive imaging techniques include \mbox{X-ray} fluorescence \cite{DAnna_Xray_fluorescence,https://doi.org/10.1002/smtd.202301610}, \mbox{X-ray} diffraction \cite{SCHULLI2018188}, angle-resolved photoemission spectroscopy (ARPES)~\cite{Procopi}, ellipsometry~\cite{Katzenmeyer2020, Young2023}, as well as scanning microwave \cite{MIM, NondestructiveAtomThin}, broadband electrostatic force \cite{bb_EFM}, and single-electron probe~\cite{doi:10.1021/acsnano.0c00736} microscopy. 
Additionally, imaging techniques based on ion-beams, such as nuclear reaction analysis~\cite{osti_5977235}, medium energy ion and Rutherford scattering~\cite{GONCHAROVA2020441,TROMBINI2019137536,SANCHEZ2011654}, can be non-destructive for conductive layers but are typically slow.
Most of these non-destructive techniques only give two-dimensional lateral information, whilst \mbox{X-ray} methods can produce three-dimensional images by tomography~\cite{Holler:2017aa,3D_tomo} at the expense of time.
Conversely, provided strong refractive index contrast, X-ray reflectometry~\cite{ZHOU1995223} can measure the vertical depth-profile of atomically thin dopant layers in a reasonably short time, of order 10~minutes per scan already since the 1990s~\cite{10.1063/1.347047}. But for thin layers with small dopant concentrations such as atomically thin layers of dopant atoms \mbox{($\delta$-layers)}, considerable modelling is required to determine the dimensions of the layers, particularly in the presence of features such as surface roughness and oxidation or other elements in the device grown on a substrate wafer.  

X-ray reflectometry can be made more sensitive to a specific element by measuring resonantly at energies around the corresponding X-ray absorption edge \cite{Macke2014}. The atomic resonance induces a large phase shift in the reflected signal, which can be used to isolate the element's contribution to the reflectivity and obtain its distribution as a function of depth. This is particularly relevant for low concentrations, as in dopant-defined devices in silicon which are based on $\delta$-layers with 3D dopant concentrations typically below 5\% \cite{Stock_As_count}. 
Here we show that for a $\delta$-layer of arsenic donors in silicon, analysis of resonant contrast \mbox{X-ray} reflectometry (RCXR) can be enormously simplified, requiring no modelling of the $\delta$-layer's host heterostructure. Furthermore, for the devices studied here SIMS gives layer thicknesses which are $>2$~nm \cite{Procopi}, considerably larger than the typically $<1$~nm thicknesses provided by measures accessing the conduction electrons in the $\delta$-layers~\cite{Stock_As_count, Procopi, Young2023}; in this work reflectometry resolves this discrepancy by showing that the chemical (arsenic) layer thicknesses are considerably smaller than the SIMS results and consistent with the electronic thicknesses established by ARPES and magneto-resistance. Therefore, RCXR is well suited for non-destructive high-throughput preliminary characterizations of semiconductor heterostructures.\\

%Intro to reflectometry and our idea
\noindent\textbf{Resonant-contrast X-ray reflectometry}

\begin{figure}  % figure setup
\centering   
    \includegraphics[width=\linewidth]{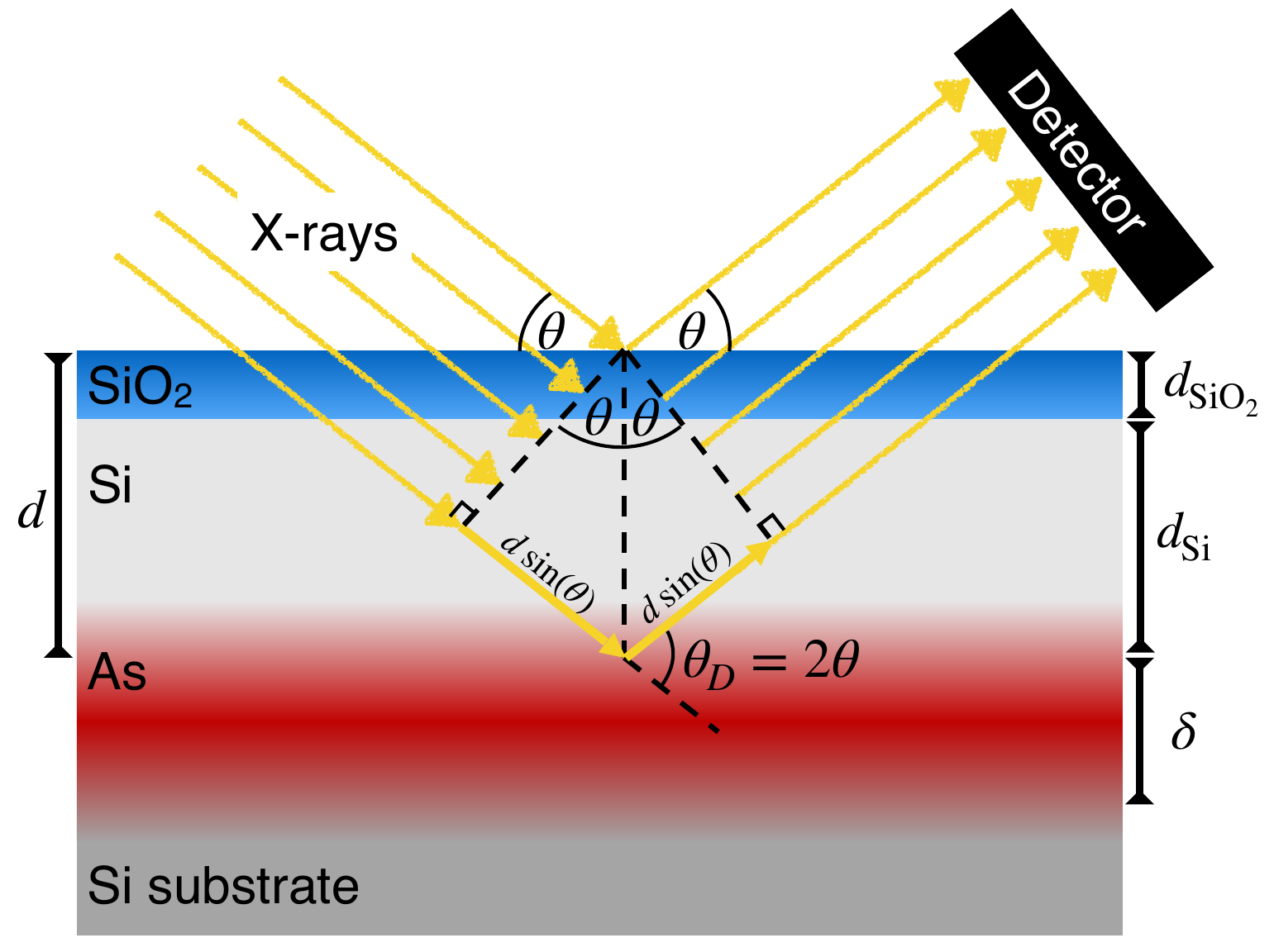} 
\caption{\textbf{X-ray reflectometry.} Schematic of the sample's layer structure and the measurement geometry. The surface consists of oxide (SiO$_2$) and silicon with a combined thickness of $d=d_{\rm SiO_2}+d_{\rm Si}$. Within the silicon lattice there is an arsenic-doped layer of thickness $\delta$.
X-rays shine on the sample with an incidence angle $\theta$ (with respect to the surface) and the specular reflection is detected by a detector placed at an angle $\theta_D=2\theta$ (with respect to the incident beam). Constructive interference occurs at $2d\sin\theta=m\lambda$ for two interfaces separated by a distance $d$, where $m$ is an integer and $\lambda$ is the X-ray wavelength.}
\label{fig_setup}
\end{figure} 

X-ray reflectometry was first used in 1954 \cite{Parratt} to measure the thickness of copper on glass and since then has become a common technique for studying a wide variety of layered materials, including surfaces, thin films and multilayers \cite{Xray_reflect_review}.
It involves measuring the specular reflection from sample surfaces (see Fig.~\ref{fig_setup}). The angle $\theta$ is swept to obtain a reflectometry measurement.

Light travelling through a medium is scattered upon changes in the medium's scattering length density 
\mbox{$\rho = r_0\sum_qN_qf_q$}, where $r_0$ is the classical electron radius, $N_q$ and $f_q=f_{1,q}+if_{2,q}$ are the number of atoms per unit volume and the complex atomic scattering factor for an atom of element $q$, respectively.
For typical semiconductors X-rays scatter weakly, hence, for a continuously varying depth profile $\rho(z)$, only single scattering events need to be considered and the measured reflection is described by the kinematic Born approximation \cite{Caticha1995,AlsNielsen_McMorrow,ZHOU1995223},
\begin{equation}
R(Q)=\frac{(4\pi)^2}{Q^2}\left|\int\rho(z)e^{-iQz}\ \mathrm{d}z \right|^2 = r_F(Q)^2|F_Q(\rho)|^2,
\label{eq_Born}
\end{equation}
where $Q=4\pi\sin\theta/\lambda$ is the $Q$-vector, $\lambda$ the X-ray wavelength, $F_Q(x)$ the Fourier transform, and \mbox{$r_F(Q)^2=\frac{(4\pi)^2}{Q^2}$} the squared Fresnel reflectivity.

The scattering length density profile $\rho(z)$ of a material is the sum of each of its elements' profiles $\sum_q\rho_q(z)$. Therefore, if an element $q$ is dilute, it can be considered to contribute perturbatively to the host profile $\rho(z)$ and, thus, to Eq.~(\ref{eq_Born}). So expanding $R(Q)$ in terms of $\rho_q(z)$, the first-order term will be due to interference between the element $q$ and the host, represented mathematically by a product of the Fourier transforms of $\rho(z)$ and the dilute elemental density profile denoted $\delta\rho(z)$ rather than the square of Fourier amplitude of $\delta\rho(z)$ appearing to second order. In the first part of our Methods section, we develop an analytic theoretical description with relevant equations that describe the consequences, with the following key conclusions:
\begin{enumerate}
\item Reflectometry data collected at a single X-ray energy can be separated into components belonging to different length-scales by Fourier filtering at the appropriate frequencies. As a consequence, if the dilute layer of element $q$ is thin compared to the other layers, its signal can readily be isolated, and if it has a Gaussian profile (of standard deviation $\sigma$) its contribution is given by:
\begin{equation}
\frac{I(Q)}{\sqrt{|F|^2_\mathrm{LF}}} = 
2r_{0}N_\text{2D}|\Delta f|\exp\left(-\frac{(Q\sigma)^2}{2}\right)\cos(Qd - \phi),
\label{eq_gauss_profile}
\end{equation}
where $d$ and $N_\text{2D}$ are the dilute layer's depth and 2D density, respectively, $\delta=2\sqrt{2}\pi\sigma$ the layer's thickness, $\Delta f = f_{q}-f_\text{host}$, $\phi = \text{arg}(\Delta f)$, $I(Q)$ the first-order term of the Born approximation's expansion in $\delta\rho(z)$ [see Eq.~(\ref{M_eq_Born_approx}) and~(\ref{M_eq_Born_interference})], and $|F|_{\mathrm{LF}}$~the low frequency part of the reflectometry.
\item Taking the difference $\Delta R(Q)$ between two reflectometry measurements at energies $E$ and $E'$ straddling a resonant edge of the dilute layer's element species, isolates the signal from the dilute layer [see Eq.~(\ref{M_eq_Born_diff1}) and (\ref{M_eq_Born_diff1_simplified})]. For a Gaussian profile this leads to Eq.~(\ref{eq_gauss_profile}) with the substitutions $I(Q) \to \Delta R(Q)/r_F(Q)^2$ and $\Delta f \to \Delta f_{E'}-\Delta f_{E}$, which directly yields the depth $d$ and thickness $\delta$ of the layer, as opposed to conventional fitting of the full reflectometry data.
\end{enumerate} 

This first-order perturbation theory has the benefit of measuring the dimensions of a dilute layer within a heterostructure of other elements without the need to model the full structure. An ideal test-bed for this are the substitutional, nm-thin dopant $\delta$-layer in silicon, with typical 2D density $N_\text{2D}\approx$10$^{14}$~cm$^{-2}$ corresponding to $N_\text{3D}=N_\text{2D}/\delta\approx$10$^{17}$~cm$^{-3}$ (for $\delta\sim$1~nm) representing $<$5\% of the total silicon atoms~\cite{Stock_As_count}.  \\
 
\begin{figure*}[th] % figure resonance analysis
\centering
\begin{adjustbox}{minipage={0.49\linewidth},valign=t}
    \subfloat[\label{fig_resonance_contrast_left}]{%
    \includegraphics[width=0.95\textwidth]{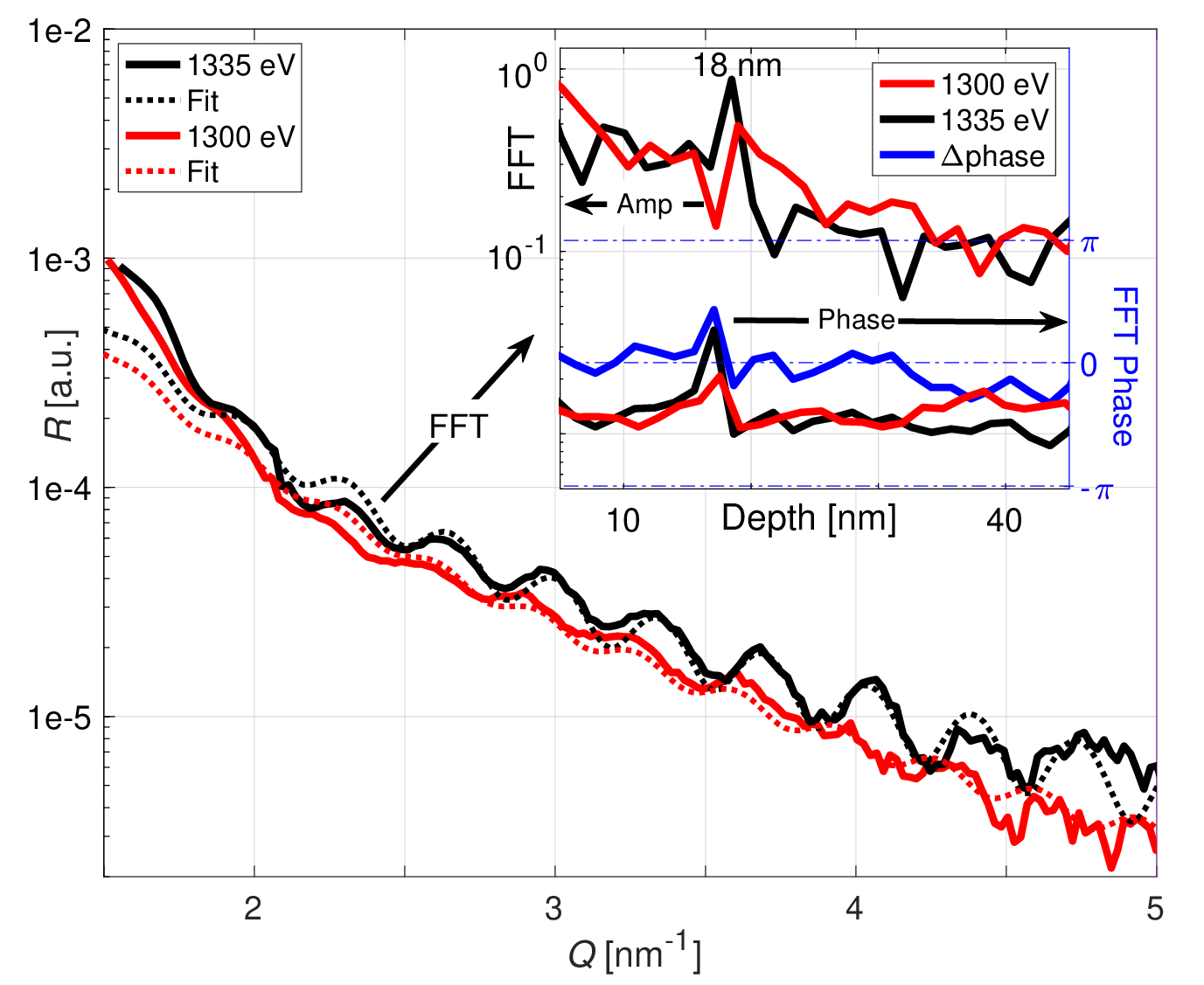}%
    }
\end{adjustbox}
\begin{adjustbox}{minipage={0.49\linewidth},valign=t} 
    \subfloat[\label{fig_resonance_contrast_right}]{%
    \includegraphics[width=0.95\textwidth]{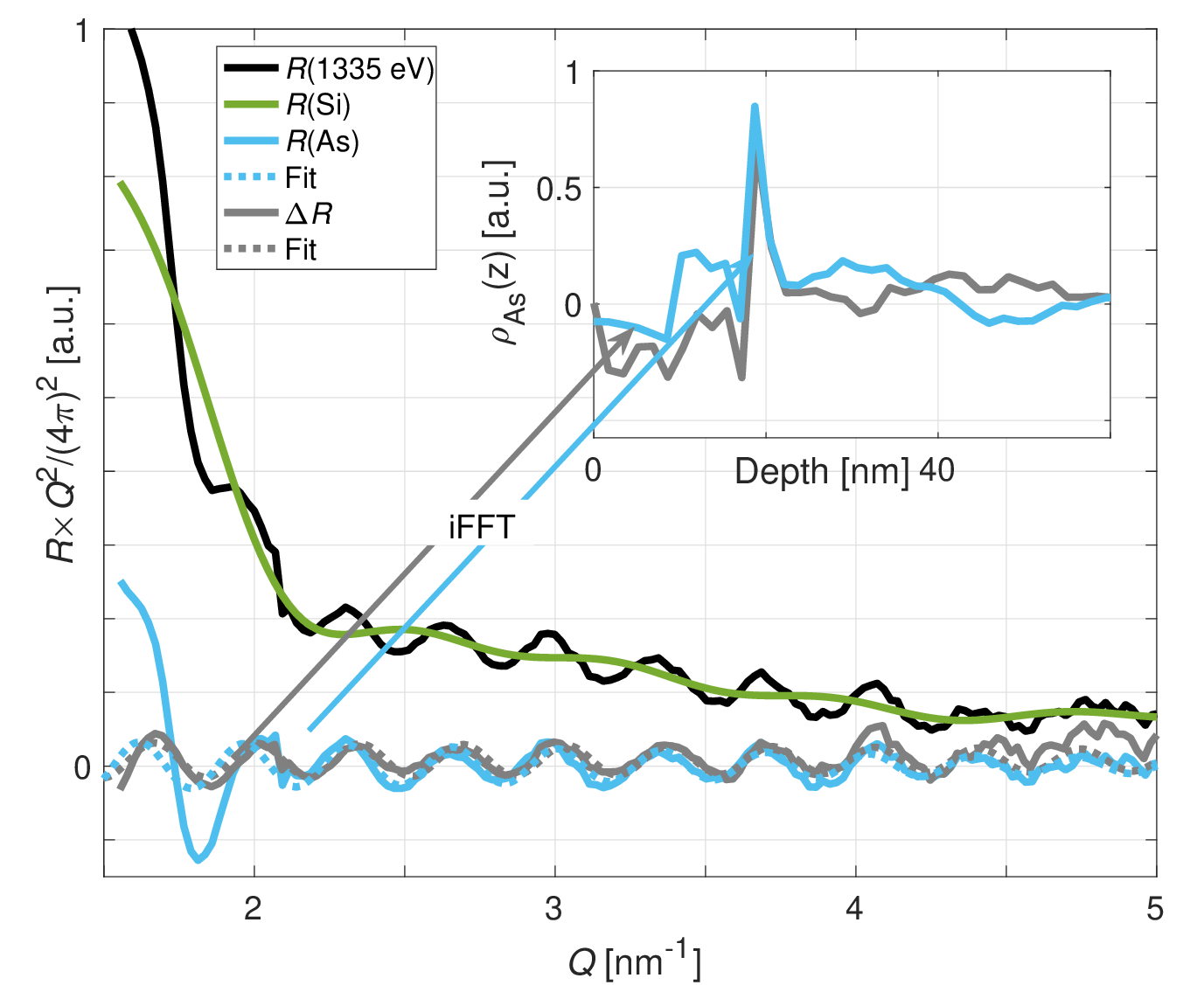}%
    } 
\end{adjustbox} 
\caption{\textbf{Dopant contribution to reflectometry across the arsenic $L_3$-edge.}  
     (a) X-ray reflectometry of sample \#1 with photon energies below (red, 1300~eV) and above the arsenic $L_3$-absorption edge (black, 1335~eV), obtained by rotating the sample and the detector to measure the specular reflection. The horizontal axis shows the wave-vector change on scattering $Q$, to directly compare the interference patterns of both energies. The inset depicts the fast Fourier transform (FFT) of the reflectometry curves on a logarithmic scale, directly yielding the depth $d$ of the arsenic $\delta$-layer. The right axis of the inset shows the FFT phase and the phase difference (blue) at both energies.
     (b) The black line shows the $R$(1335~eV) data from (a) multiplied by $Q^2$. A Fourier transform filter was used to separate frequencies corresponding to reflections from depths less than 10~nm ($R$(Si), green) and greater than 10~nm ($R$(As), blue). The gray line represents the difference $\Delta R$ between the reflectometry measured above (1335~eV) and below (1300~eV) the arsenic $L_3$-edge. Both $R$(As) and $\Delta R$ have been divided by $\sqrt{R(\text{Si})}$ to isolate the arsenic contribution. The inset shows the arsenic scattering length density $\rho_\text{As}$, obtained from $R$(As) and $\Delta R$ as described by Eq.~(\ref{M_inverse_transform}) in the Methods. Dotted curves denote fits using the DYNA program and Eq.~(\ref{eq_gauss_profile}) in (a) and (b), respectively.}
\label{fig_resonance_contrast}
\end{figure*}

\noindent\textbf{Extracting the arsenic signal from reflectometry}

The samples used to demonstrate our method are Si(001) wafers with arsenic $\delta$-layers (Si:As) at varying depths $d=15 - 75$~nm below the surface and an oxide (SiO$_2$) surface layer of width \mbox{$d_{\rm SiO_2}\approx1$~nm}~\cite{Al-Bayati1991} (see Fig.~\ref{fig_setup}). They were prepared as described in the Methods, with resulting two-dimensional dopant density \mbox{$N_{\mathrm{2D}}\approx10^{14}$~cm$^{-2}$}, as determined by SIMS, STM, magneto-resistance, and X-ray fluorescence~\cite{Stock_As_count, DAnna_Xray_fluorescence}. 
In addition, one control sample features a buried oxide layer instead of arsenic.

In the doped region, considering a \mbox{$d_{\rm As}\approx1$~nm} thick layer, $<5\%$ of silicon atoms are replaced by arsenic, while the oxide contains twice as many oxygen~atoms as silicon. Therefore, the optical contrast between the silicon and Si:As is small compared to that between oxide and silicon. 
Figure~\ref{fig_resonance_contrast_left} shows typical reflectometry data for our samples.
The black data are for an energy of 1335~eV, above the arsenic \mbox{$L_3$-absorption} edge at 1324~eV \cite{HENKE1993181}, and the red ones for 1300~eV, below the edge. Both datasets contain distinct fast oscillations due to interference between reflections from the $\delta$-layer and those from the surface region. The periodicity of the oscillations is related to the dopant layer depth $d$ through the Bragg condition $d=m\lambda/2\sin\theta$ (corrections from the refractive index $n$ can be omitted for X-rays), \textit{i.e.}, a smaller $d$ implies a larger period. In the inset of Fig.~\ref{fig_resonance_contrast_left}, a depth of $d=18$~nm is obtained from the Fourier transform. 
The fast oscillations are modulated by an envelope that varies slowly with $Q$ on a scale inversely proportional to the thin arsenic layer thickness $\delta$ and broadened also by the surface roughness.

At very small angles (\textit{i.e.} for $Q<1.5$~nm$^{-1}$) the signal contains no meaningful structure on account of a combination of total external reflection and the leakage of light directly to the detector due to growth of the beam footprint to larger than the sample. At higher angles we can analyze the data quantitatively: the oscillations decay because the overall signal intensity decreases by the factors $r_F(Q)^2\propto Q^{-2}$ and $H(Q)$ [see Eq.~(\ref{M_eq_sym_interference}) in Methods]. Surface roughness reduces the intensity by an additional factor which can be modelled by $e^{-Q^2\sigma_r^2}$, where $\sigma_r$ is the root mean square roughness~\cite{nevot1980characterization,AlsNielsen_McMorrow}. This roughness was measured with STM and atomic force microscopy (AFM, see Methods), and found to be 0.1~nm, consistent with other studies~\cite{Goh2009}.

\begin{figure*} % figure resonance
\centering
\begin{adjustbox}{minipage={0.49\linewidth},valign=t}
    \subfloat[\label{fig_resonance_left}]{%
    \includegraphics[width=0.95\textwidth]{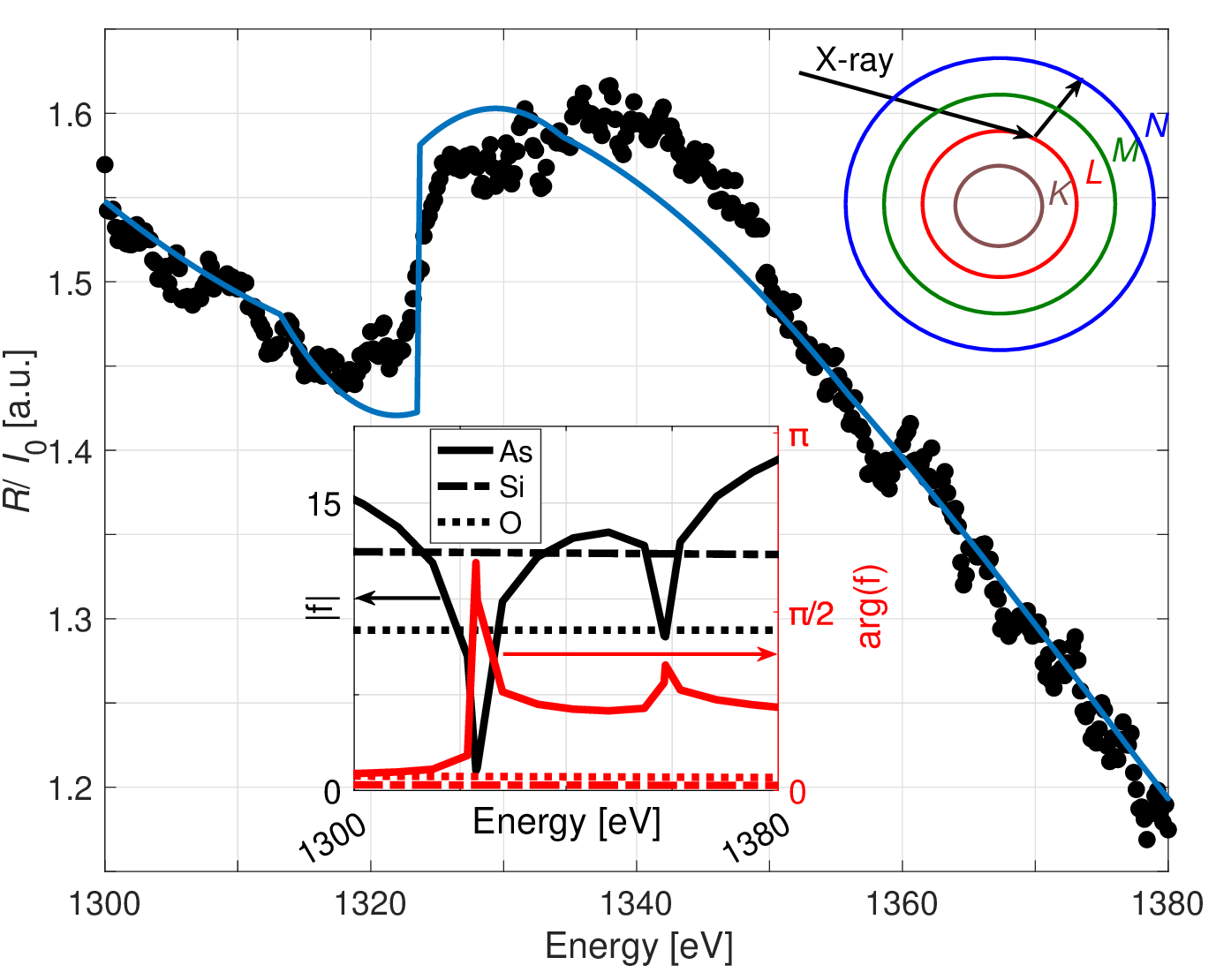}%
    }
\end{adjustbox}
\begin{adjustbox}{minipage={0.49\linewidth},valign=t} 
    \subfloat[\label{fig_resonance_right}]{%
    \includegraphics[width=0.95\textwidth]{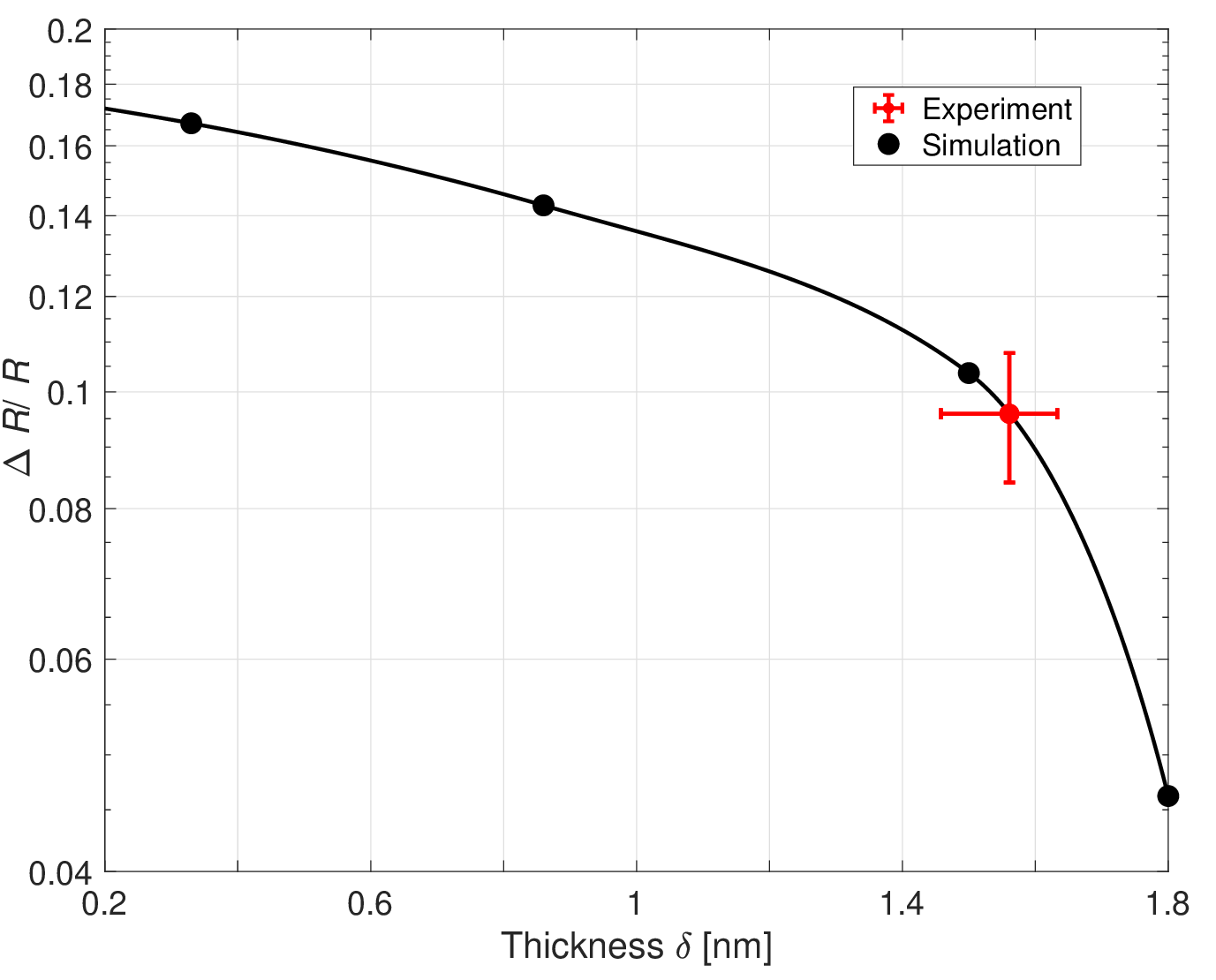}%
    }
\end{adjustbox}   
\caption{\textbf{Resonant reflectometry across the arsenic $L_{2,3}$-edges.} (a) Sharp reflectometry edge at the arsenic $L_3$-absorption edge resonance of 1324~eV in sample \#1, at an incidence angle of $\theta=10^{\circ}$, normalised by the incident photon flux $I_0$. The blue line is a fit assuming $\delta$ = 1.5~nm (see Methods and Fig.~\ref{Spectrum_fits}). The top inset depicts the resonance process where an X-ray photon is absorbed and excites an electron from the $L$-shell to the unoccupied $M$- or $N$-shells. The bottom inset depicts the energy dependence of the arsenic, silicon, and oxygen scattering factors in full, dashed, and dotted lines, respectively \cite{NIST_table}. Only the arsenic scattering factor changes noticeably in this energy range.
(b) Simulated (black) and experimental (red) relative reflection change~$\Delta R/R$, assuming $N_\text{2D} = 2.77\times10^{14}$~cm$^{-2}$. The black line is a guide to the eye used to estimate the error of the arsenic layer thickness $\delta$ (see Methods). The intersection of the data and the simulation gives $\delta = 1.6 \pm0.5$~nm for sample~\#1.}
\label{fig_resonance}
\end{figure*}

The phase between the two signals, shown in the inset of Fig.~\ref{fig_resonance_contrast_left} is shifted at the arsenic resonance edge, as described by Eq.~(\ref{M_eq_Born_approx}), where the second term is phase sensitive and proportional to the arsenic scattering length density. The expected phase shift is given by the scattering factor change $\Delta f=f_\text{As}-f_\text{Si}$, as shown in equation Eq.~(\ref{M_eq_Born_interference}). Thus, reflectometry data taken at 1300~eV and 1335~eV will be shifted by \mbox{$\text{arg}\left[\Delta f(1335\text{eV})\right]-\text{arg}\left[\Delta f(1300\text{eV})\right]\approx (0.33\pm0.10)\pi$,} using scattering factors from \cite{NIST_table}. This is in agreement with the measured phase shift of $(0.27\pm0.03)\pi$ at 18~nm indicated by the peak in the inset of Fig.~\ref{fig_resonance_contrast_left}.

Conventional fits to the data using the DYNA~\cite{Dyna} software in Fig.~\ref{fig_resonance_contrast_left} are shown with dotted lines, where the three-dimensional density $N_\text{3D}$ and the $\delta$-layer thickness are obtained from fitting the resonance in Fig.~\ref{fig_resonance}, to reduce the number of free fitting parameters. Direct fits to the data for a single photon energy are unable to give reliable results for the \mbox{$\delta$-layer} thickness $\delta$, due to the large fitting parameter space containing each layer's thickness, density, and roughness metrics, highlighting the need for a dedicated method to extract $\delta$.

To analyze the data at an X-ray energy of 1335~eV, above the arsenic $L_3$-edge (black data in Fig.~\ref{fig_resonance_contrast_left}), the data are multiplied by $Q^2/(4\pi)^2$ to remove the $Q^{-2}$~divergence expected from Eq.~(\ref{eq_Born}). Thereafter, a Fourier transform filter is used to separate oscillations corresponding to reflections from depths less than 10~nm that we attribute to the silicon layer ($R$(Si), shown in green in Fig.~\ref{fig_resonance_contrast_right}), and oscillations that stem from depths greater than 10~nm, which after division by $\sqrt{R(\text{Si})}$ we attribute to the arsenic layer ($R$(As), shown in blue in Fig.~\ref{fig_resonance_contrast_right}).
Under the assumption that the arsenic layer has a Gaussian profile, $R$(As) can be fitted to Eq.~(\ref{eq_gauss_profile}) (dotted blue line, Fig.~\ref{fig_resonance_contrast_right}) to obtain the arsenic layer depth and thickness, here $d = 18.1\pm0.1$~nm and $\delta=0.9\pm0.2$~nm, respectively. % ($\delta=$0.9 (0.7\text{--}1.1)~nm)

In the Methods section, we show that for a thin arsenic layer, the slow $Q$ oscillations (see Fig.~\ref{fig_resonance_contrast_right}, green) are largely due to the amplitude $|F_Q(\rho)|$ for the unperturbed host. Thus, we obtain the inverse Fourier transform of the arsenic factor $F_Q(\delta{\rho}(z))$ by first subtracting $|F_Q(\rho)|$ from the raw data divided by $Q^{2}/(4\pi)^{2}$ and then dividing by $\sqrt{|F_Q(\rho)|}$ (see Fig.~\ref{fig_resonance_contrast_right}, blue). Finally, taking the Fourier transform of $|F_Q(\delta{\rho}(z))|$ results in the scattering length density profile $\rho_\text{As}(z)$ [Eq.~(\ref{M_inverse_transform})] shown in the inset of Fig.~\ref{fig_resonance_contrast_right}, with a large density at the expected arsenic layer depth. The resolution in $z$ is $\sim$1.5~nm and dictated by the sampling range in $Q$; too low to resolve the shape of a $\delta<1$~nm dopant-layer profile.\\ 

% resonant contrast
\noindent\textbf{RCXR isolation of the $\delta$-layer signal}

To quantify the modulation of the fast oscillations in the reflectivity, we look at the difference measured below and above the arsenic $L_3$-edge resonance energy at 1300~eV and 1335~eV, respectively, similarly to how magnetism can be extracted from the difference between left- and right-handed circularly polarized \mbox{X-rays}~\cite{doi:10.1021,PhysRevLett.131.036201}. 

Figure~\ref{fig_resonance_contrast_right} shows in gray the difference $\Delta R$ of the reflectivities at the two energies divided by square root of the silicon contribution. 
Since the quantity of interest is the Fourier transform of the density profile $\rho(z)$, the data were multiplied by $Q^2/(4\pi)^2$ consistent with Eq.~(\ref{M_eq_Born_approx}).
The subtraction removes the zeroth-order term due to the oxide and silicon layers and keeps the first-order term due to the arsenic-doped layer, because the scattering factors of the former are almost constant in this energy range, while they change drastically for the latter (see inset of Fig.~\ref{fig_resonance_right}). The division by the Si/SiO$_2$ contribution leaves a fast oscillation, with a period corresponding to the arsenic \mbox{$\delta$-layer} depth, modulated by a slow envelope with a decay corresponding to the arsenic $\delta$-layer thickness (described by Eq.~(\ref{M_eq_Born_diff1})). The same slow modulation of the fast oscillation is present in the data for $R(\text{As})$ in Fig.~\ref{fig_resonance_contrast_right}. 
Assuming a Gaussian arsenic density profile, the data are fitted to Eq.~(\ref{eq_gauss_profile}) yielding a thickness of $\delta=0.6_{-0.2}^{+0.7}$ ($0.4 -1.3$)~nm and depth of $d=18.4\pm0.1$~nm in agreement with the result from a single energy (see Fig.~\ref{fig_resonance_contrast_right}, gray), where $\delta$ is taken to be the full width at half maximum of the Gaussian profile and the uncertainty indicates the fit's 95\% confidence interval.
We emphasise that the value extracted here represents an upper bound on the arsenic layer thickness due to the high-angle cutoff due to noise and interface roughness~\cite{HENKE1993181,Roughness_comp,doi:10.1063/1.1332095}. 
Surface roughness contributions add in quadrature to the result, here roughness was found to be of the order of~0.1~nm by STM and AFM measurements (see Methods) and is, therefore, negligible.

The Fourier transform of the $\Delta R$ data in Fig.~\ref{fig_resonance_contrast_right} also leads to the arsenic profile $\rho_\text{As}(z)$ (see inset of Fig.~\ref{fig_resonance_contrast_right}).
The obtained profile is in good agreement with that obtained from a single scan of the reflectivity, suggesting that our Fourier filtering procedures for single X-ray energies, not required when we are taking differences between data collected at different X-ray energies, are adequate. Furthermore, the resonance subtraction method suppresses fluctuations on the sides of the $\delta$-layer peak.  

\begin{figure*} % figure d delta dimensions
\centering
\begin{adjustbox}{minipage={0.49\linewidth},valign=t}
    \subfloat[\label{fig_d_delta_left}]{%
    \includegraphics[width=0.95\textwidth]{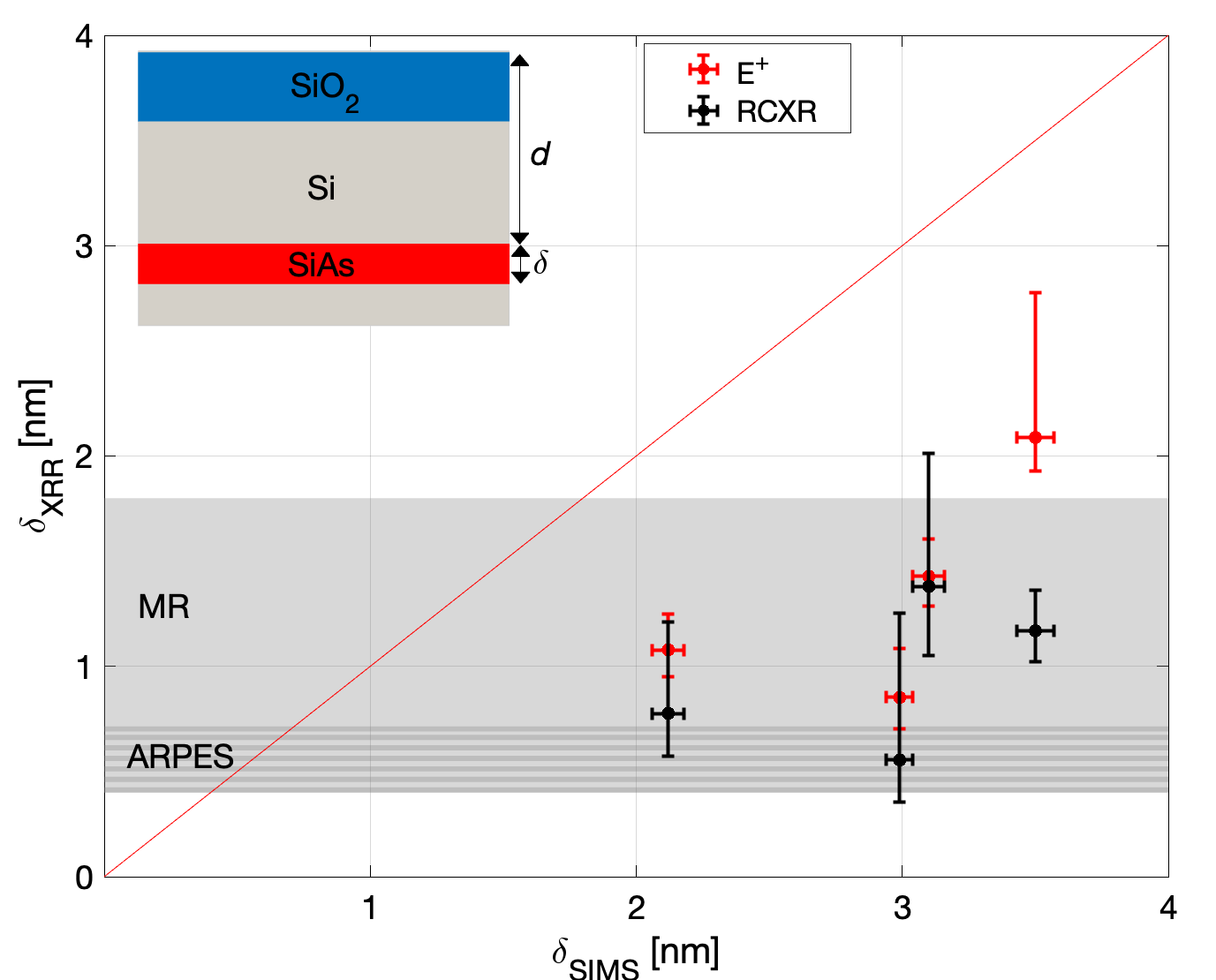}%
    }
\end{adjustbox}
\begin{adjustbox}{minipage={0.49\linewidth},valign=t}  
    \subfloat[\label{fig_d_delta_right}]{%
    \includegraphics[width=0.95\textwidth]{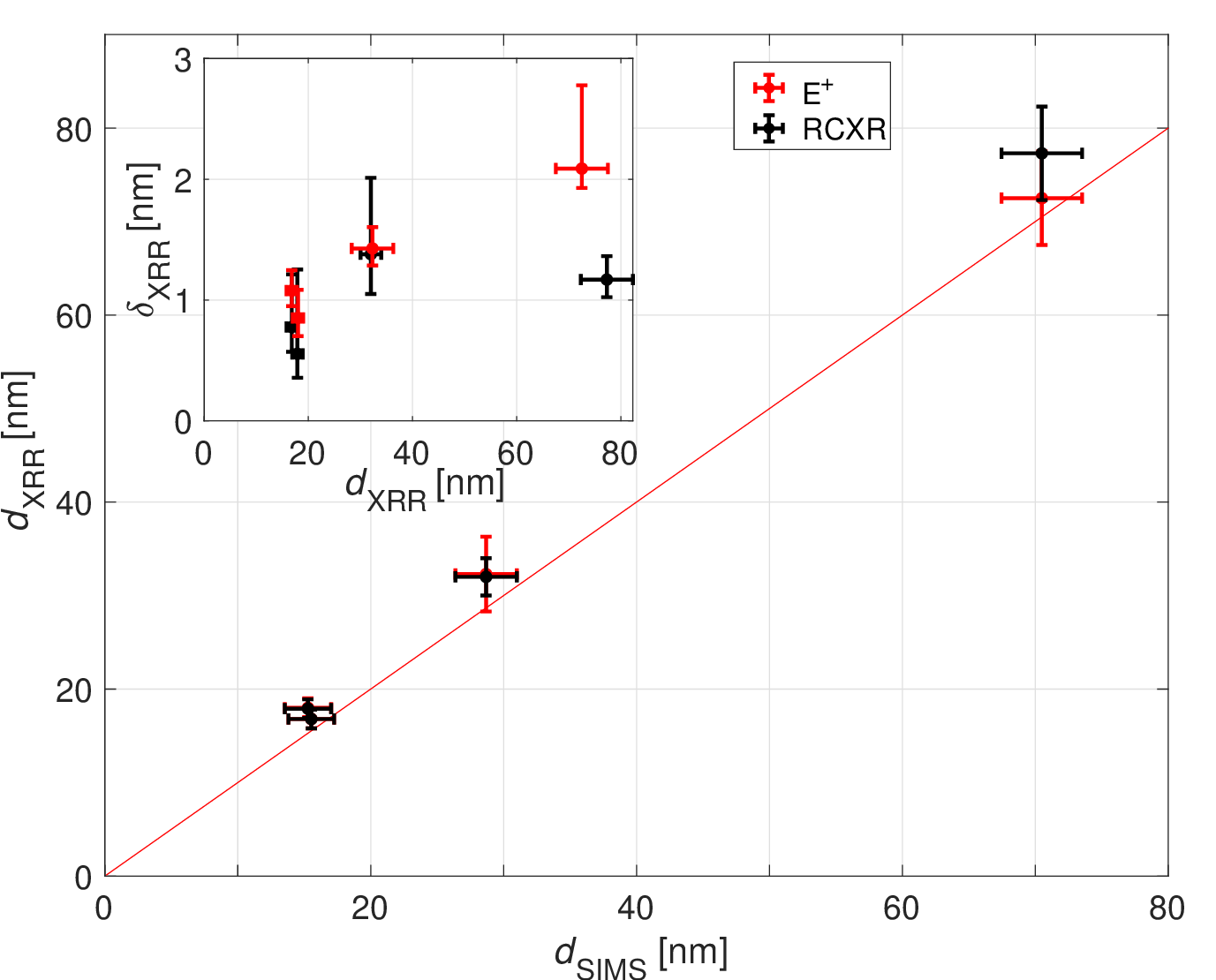}%
    }
\end{adjustbox} 
\caption{\textbf{Comparison of dopant layer thicknesses and depths.} The thickness $\delta$ (a) and depth $d$ (b) of the Si:As layer below the surface, obtained by RCXR (black) and a single energy above the $L_3$-edge (E$^+$, red), as well as SIMS are plotted on the vertical and horizontal axes, respectively.
The inset in panel (b) shows the dependence of $\delta$ on $d$ as measured with X-ray reflectometry (XRR).
The shaded regions in (a) denote the thickness range obtained by ARPES and magneto-resistance~(MR) of samples prepared in the same way.
SIMS depth errors are from the uncertainty in the sputter rate (see Methods). XRR~errors for E$^+$ and RCXR are taken from the 95\% confidence interval of fits to Eq.~(\ref{eq_gauss_profile}).}
\label{fig_d_delta}
\end{figure*}

RCXR experiments were performed on five different samples, of which four contained Si:As $\delta$-layers and one reference sample a buried oxide layer. They were also measured with SIMS (see Methods).
The results for both techniques are shown in Fig.~\ref{fig_d_delta}, where the data in red are obtained from Fourier filtered data at a single energy above the arsenic $L_3$-edge and the data in black from the resonant contrast method. 
%The results for the thickness $\delta$ rely on the assumption that the dopant layer depth is larger than its thickness and that of other layers, as well as that the layer has a Gaussian profile.

Our isolation of the $\delta$-layer signal based on fitting the amplitude modulation of the dopant layer has the advantage that it relies neither on extensive multi-variable fitting, nor on prior knowledge or hypotheses concerning the sample composition other than assuming Gaussian profiles. Additionally, for thin layers, the amplitude modulation occurs over a large $Q$-range, such that the estimation of the layer thickness has high accuracy.\\

\noindent\textbf{Arsenic $L_{2,3}$-edge resonance measurements} % resonant edge analysis

The previous section shows that upper bounds on \mbox{$\delta$-layer} thicknesses are readily obtained from examination of the rapid oscillations associated with their depth. On the other hand, short wavelength disorder can introduce noise at high momentum transfers $Q$ and, thereby, complicate the determination of lower bounds of $\delta$. 
However, it is possible to obtain alternative thickness estimates by relying on photon energy scans through the dopant resonance at low $Q$, provided that the two-dimensional dopant density $N_\text{2D}$ is known precisely. 
The \mbox{arsenic $L_{2,3}$-edge} resonances are visible when measuring the reflected intensity as a function of X-ray energy at a fixed angle (see Fig.~\ref{fig_resonance_left}), where the arsenic complex atomic scattering factor changes abruptly, while the silicon and oxygen atomic scattering factors are smooth functions of energy (see inset of Fig.~\ref{fig_resonance_left})~\cite{NIST_table}.
The scattering length density of the layers and the resonance intensity are determined by the effective three-dimensional dopant density $N_\text{3D} = N_\text{2D}/\delta$ and the Born equation Eq.~(\ref{eq_Born}), such that the thinner the dopant layer ($\delta$), the stronger the resonance (because $N_\text{3D}$ is larger) at any non-zero $Q$ (described by Eq.~(\ref{M_eq_Born_interference})).
The relative change in reflection, \textit{i.e.}, \mbox{$\Delta R/R = [R(1330~\text{eV}) - R(1320~\text{eV})]/R(1320~\text{eV}),$} depends mostly on $N_\text{3D}$ of the layer that undergoes the resonance, here the arsenic $\delta$-layer, but its absolute value also hinges on the characteristics of all other layers [apparent from Eq.~(\ref{M_eq_Born_approx}), see Methods]. 
As a consequence, if~$N_\text{2D}$ and the layer depth are known, it is possible to obtain its thickness $\delta$ directly from the resonance spectrum.

The two-dimensional dopant density $N_\text{2D}$ is readily obtained by X-ray fluorescence \cite{DAnna_Xray_fluorescence}.
In principle, it is straightforward to record the X-ray fluorescence and reflectivity simultaneously, as the fluorescence photons have an isotropic distribution. $N_\text{2D}$ was measured for one sample in this work (sample \#1 shown in Fig.~\ref{fig_resonance_contrast}~\&~\ref{fig_resonance}), and was found to be $N_\text{2D} = (2.77\pm0.14)\times10^{14}$~cm$^{-2}$ by the same method as in~\cite{DAnna_Xray_fluorescence}.
The relative change in reflection $\Delta R/R$ at the arsenic $L_{2,3}$-edges as a function of the dopant layer thickness was also calculated with the DYNA program \cite{Dyna}  (see Fig.~\ref{fig_resonance_right}, black).
The layer depth used for this calculation was obtained in Fig.~\ref{fig_resonance_contrast_left}.
As expected, decreasing the dopant layer thickness  and/or increasing~$N_\text{3D}$ at fixed $N_\text{2D}$, increases the relative change in resonance intensity $\Delta R/R$.
In Fig.~\ref{fig_resonance_right}, the experimental value of $0.09\pm0.01$ for $\Delta R/R$ extracted from Fig.~\ref{fig_resonance_left} is shown in red. An arsenic $\delta$-layer thickness of $1.6\pm0.5$~nm is deduced. Within errors this agrees with the value from the previous section ($\delta=0.6_{-0.2}^{+0.7}$~nm), but relies on fitting the resonance with each layer's thickness, density, and roughness, resulting in a large number of fitting parameters, and prior knowledge about all layers in the sample.\\

% results/discussion
\noindent\textbf{Discussion} 

RCXR and SIMS experiments were performed on four samples containing an arsenic $\delta$-layer. 
Figure~\ref{fig_d_delta} shows that the depths $d$ of the shallower samples agree within error bars, whereas SIMS underestimates the depth of deeper Si:As $\delta$-layers. This discrepancy might originate from the variability of the SIMS sputter rate during a measurement \cite{SIMS_sputter}.
The $\delta$-layer thickness (see Fig.~\ref{fig_d_delta_left}) is lower when measured with X-ray reflectometry, as is expected since the SIMS resolution is $\approx2$~nm \cite{PhysRevB.101.245419}.
The values for the layer thickness measured with RCXR denote an upper bound, in our case determined by the maximum momentum transfer $Q_{\rm max}$=5~nm$^{-1}$. In the present experiment where $\lambda$=0.94~nm, $Q_{\rm max}$ is fixed by $\theta_{\rm max}=22^{\circ}$, the maximum angle of incidence, which can easily be raised in experiments with  experiments with next-generation synchrotron sources and instruments. Nevertheless, the RCXR results show that our arsenic $\delta$-layer samples are as thin as~0.6~nm. Also, the upper-bound thicknesses measured here are in good agreement with results from ARPES \cite{Procopi}, where Si:As samples fabricated in the same way were measured to be~0.4 to 0.7~nm thick (see shaded area in Fig.~\ref{fig_d_delta_left}) from the point of view of the 2D electron liquids hosted by the $\delta$-layers. Additionally, one of the samples in this study ($\delta_{RCXR} = 1.4\pm0.5$~nm) was etched into a Hall bar geometry and contacted for low-temperature magneto-resistance \cite{NIST_SiP}, from which a conductive layer thickness of~$0.97\pm0.02$~nm was extracted \cite{DAnna_Xray_fluorescence}. Magneto-resistance of ten equivalent arsenic $\delta$-layers yielded a thickness range from 0.4~nm to 1.8~nm (see shaded area in Fig.~\ref{fig_d_delta_left}), further corroborating our method and results.
 
Dopant $\delta$-layers in silicon have been extensively studied with magneto-resistance \cite{NIST_SiP}, SIMS \cite{Wang2018}, and ARPES~\cite{Procopi}. The agreement between data in Fig.~\ref{fig_d_delta} measured at a single energy above the arsenic $L_3$-edge (red) and by RCXR~(black) shows that RCXR, treated within the first-order perturbation theory described here, is able to extract the arsenic dopant layer depth and thickness reliably without extensive structural modelling. The fact that the difference in phase shift of the oscillations on and off resonance matches the tabulated atomic phase shifts certifies the applicability of perturbation theory. 

While RCXR is used here to enhance the sensitivity to a single two-dimensional dopant layer in silicon, naturally it also increases the sensitivity to specific layers in any multilayer material, including group IV and III-V semiconductors, as well as quantum materials where electron correlations are strong. For example, owing to the strong soft X-ray resonances of the transition metal \mbox{$L$-edges} and the oxygen \textit{K}-edge, RCXR is ideally suited to non-destructively characterize oxide heterostructures \cite{keimer2015quantum,doi:10.1126/science.1178863}. An ideal venue for our technique are the recently discovered superconducting infinite-layer nickelates \cite{li2019superconductivity,doi:10.1126/sciadv.abl9927}, for which the exact structure of the infinite-layer phase and other topotactically related phases are currently a matter of discussion \cite{10.1063/5.0005103,Nomura_2022}. In particular, there is the possibility of an oxygen-ordered impurity superlattice \cite{parzyck2023absence}, for which RCXR can be used to isolate and amplify the oxygen signal.
 
% conclusion
In conclusion, X-ray reflectometry can be made sensitive to specific elements in layered samples by performing a differential measurement above and below a resonance absorption edge of the respective element. We show that with this technique and a simple perturbation expansion of the differential reflectivity, it is possible to isolate the signal from one specific element without needing to model the material, yielding an upper bound as low as 0.9~nm for the thickness of arsenic-doped $\delta$-layer buried in silicon with a 3D arsenic density $<5\%$ of the total silicon density. In principle, the upper bound can be extended by increasing the X-ray fluence to scan to higher reflection angles, as well as by reducing the sample surface roughness.
Combination of this approach with nano X-ray fluorescence detection~\cite{DAnna_Xray_fluorescence} will make it possible to non-destructively characterize samples in three-dimensions in a single experiment. Finally, specular and off-specular RCXR with nano-X-ray beams featuring spot sizes of less than 10~nm \cite{ID16A} will enable imaging of patterned dopant structures such as the gates, sources and drains needed for future classical and quantum electronics.
\\

\noindent\textbf{Methods}

\noindent\textbf{First-order resonant-reflectometry theory.}
For a single interface between two materials, the reflectance $R$ for $\sigma$ polarised light, is given by the Fresnel equation
\begin{equation}
R = \left|\frac{n_1\sin\theta_i-n_2\sin\theta_t}{n_1\sin\theta_i+n_2\sin\theta_t}\right|^2,
\label{M_eq_Fresnel}
\end{equation}
where $\theta_i$ and $\theta_t$ are the beam incidence and transmission angles and are equal ($\theta_i = \theta_t = \theta$) for specular reflection as defined in Fig.~\ref{fig_setup}. $n_1$ and $n_2$ are the refractive indices of the top and bottom layer, respectively. 
The reflectance~$R$ is related to the material's atomic scattering factors $f_1$ and $f_2$ through the refractive index \cite{HENKE1993181,NIST_table} 
\begin{equation}
n=1-\frac{r_0}{2\pi}\lambda^2\sum_qN_qf_q = 1-\frac{\lambda^2}{2\pi}\rho,
\label{M_eq_ref_index}
\end{equation}
where $r_0$ is the classical electron radius, $\lambda$ the photon wavelength, $N_q$ and $f_q=f_{1,q}+if_{2,q}$ are the number of atoms per unit volume and the complex atomic scattering factor for an atom of element $q$, respectively. $\rho$ is the scattering length density. $f_1$ as well as $f_2$, and hence also $\rho$, depend on the incident beam energy.

For multiple layers, the reflections at boundaries interfere. Depending on the incident beam angle $\theta$ and the layer thicknesses, this interference can be destructive or constructive. For two interfaces, as shown in Fig.~\ref{fig_setup}, the condition for constructive interference is given by Bragg's law $2d\sin\theta=m\lambda$, and the reflectance is periodic in the wave-vector change on reflection $Q=4\pi\sin\theta/\lambda$. 
%For three interfaces, the interference is modulated by a cardinal sine (sinc), analogous to the double-slit experiment. 

In most cases, rather than consisting of perfectly homogeneous and sharp layers, samples have a continuously varying scattering length density profile $\rho(z)$. If scattering is weak, multiple scattering events can be neglected, and the reflected signal is simply the sum of the partial waves emanating from the different scatterers driven by the unperturbed incident field. The resulting far-field amplitude is given by the kinematic Born approximation~\cite{Caticha1995,AlsNielsen_McMorrow,ZHOU1995223}\footnote{The sign in the exponential in Eq.~(\ref{M_eq_Born}) depends on the convention for the imaginary part of $\rho$. We follow Chantler \cite{Chantler2000}, corresponding to the form of the Born approximation in Caticha~\cite{Caticha1995}.},
\begin{equation}
R(Q)=\frac{(4\pi)^2}{Q^2}\left|\int\rho(z)e^{-iQz}\ \mathrm{d}z \right|^2 = r_F(Q)^2|F_Q(\rho)|^2,
\label{M_eq_Born}
\end{equation}
which is essentially the squared amplitude of the Fourier transform $F_Q(x)$ of $\rho(z)$ for a given $Q$-vector times the squared Fresnel reflectivity $r_F(Q)^2=\frac{(4\pi)^2}{Q^2}$. %, where $Q$ depends on the beam energy and incidence angle. 
For a continuously varying density profile, the spatial resolution of reflectometry is of the order of the inverse of the greatest $Q$-vector, which is $\approx 0.2$~nm for the 1300~eV X-rays used in this work.

Data collected with X-ray reflectometry are commonly analysed using software that solves Maxwell's equations throughout the material \cite{GenX,Dyna}. To obtain good fits it is necessary to consider each layer's atomic-species, density, thickness and roughness, resulting in a high number of fitting parameters. 
However, often one is faced with the problem of characterizing a layer with low contrast hosted by an otherwise known heterostructure, such as nano-electronic stacks involving III-V or group IV semiconductors, where dopant structures are typically buried, such as the silicon and oxide surface layers above our dopant $\delta$-layers. 

We introduce a new method to find and isolate the contribution to X-ray reflection from such a thin layer. First, we consider the general problem of a heterostructure with a scattering length density profile~$\rho(z)$, to which we add a second scattering length density profile $\delta\rho(z)$, accounting for replacement of atoms (\textit{e.g}. As for silicon or aluminum for arsenic in III-V devices) from the original heterostructure. The Born approximation, expanding the square at the right of Eq.~(\ref{M_eq_Born}), yields three terms in the reflectivity of the perturbed heterostructure:
\begin{equation}
\frac{R(Q)}{r_F(Q)^2} = |F_Q(\rho)|^2 + \underbrace{2\Re\left[F_Q(\rho)\bar{F_Q}(\delta\rho)\right]}_{I(Q)} + |F_Q(\delta\rho)|^2.
\label{M_eq_Born_approx}
\end{equation}

Eq.~(\ref{M_eq_Born_approx}) contains a background from the unperturbed heterostructure, an interference term $I(Q)$ which is first order in $\delta\rho(z)$, and a signal, of second order in $\delta\rho(z)$, from the perturbation. For a weak perturbation from a small number of atoms, we can ignore the second-order term, but the zeroth-order background remains troublesome as the major contributor to the reflectivity. However, if $\delta\rho(z)$ could be modulated by a known prefactor without affecting $\rho(z)$, differences of reflectivities for different prefactors would remove the unperturbed signal, leaving simply the interference term to analyze. The tunable photon energy of a synchrotron source provides a natural modulator for scattering lengths, especially near resonances which are unique fingerprints for particular elements. If the perturbation entails simply replacing atoms of species A with species B, then 
\begin{equation}
\delta\rho(z)= r_0\Delta fN_\text{2D}\delta\tilde{\rho}(z),
\label{M_eq_ansatz}
\end{equation}
where $\Delta f = f_B-f_A$, $N_\text{2D}\delta\tilde{\rho}(z)$ is the 3D density of A atoms at depth $z$ replaced by B atoms, $N_\text{2D}$ is the 2D density of the layer, and the integral over $z$ of $\delta\tilde{\rho}(z)$ is unity. This integral is also the long wavelength ($Q\to 0$) limit of the Fourier transform of $\delta\tilde{\rho}(z)$, so that $N_\text{2D}$ sets the scale of the interference term in Eq.~(\ref{M_eq_Born_approx}), which we can rewrite as 
\begin{equation}
I(Q) = 2r_{0}N_\text{2D} \Re\left[\bar{\Delta f}F_Q(\rho)\bar{F_Q}(\delta\tilde{\rho})\right].
\label{M_eq_Born_interference}
\end{equation}

To progress, we assume that for a relatively deep \mbox{$\delta$-layer} at depth $d$, $\delta\tilde{\rho}(z)=h(z-d)$, where $h(z)$ is an even function. Hence \mbox{$F_Q(\delta\tilde{\rho})= \exp(-iQd) H(Q)$} where $H(Q)$ is real, giving
\begin{equation}
I(Q) = 2r_0N_\text{2D}H(Q)\Re\left[\bar{\Delta f}\exp(iQd) F_Q(\rho)\right].
\label{M_eq_sym_interference}
\end{equation}

If $d$ is the largest characteristic length in the problem, \textit{i.e.} $d$ is much larger than the thickness of a probable surface oxide layer which contributes to $F_Q(\rho)$, Eq.~(\ref{M_eq_sym_interference}) gives several results: 
\begin{enumerate}

    \item There are rapid oscillations with period $2\pi/d$ and a phase fixed largely by the complex scattering amplitude \mbox{$\Delta f$}, implying that changes in photon energy will change the phase of the oscillations. On the other hand, changes in $N_\text{2D}$ or $H(Q)$ (as long as $h(z)$ remains even) will not change this phase. Furthermore, the third term in Eq.~(\ref{M_eq_Born_approx}) scales as $|\Delta f|^2$ and, therefore, does not undergo a phase shift on changing photon energy. This means that the importance of the interference term in Eq.~(\ref{M_eq_Born_approx}) and the validity of first-order perturbation theory is readily checked by examining the phase shift. 

    \item Oscillation amplitudes scale by \mbox{$N_\text{2D}\left|\Delta fF_Q(\rho)\right|H(Q)$}, meaning that if we operate near a resonance of the substitute atom B but far from resonances of the unperturbed heterostructure, then there will be an anomaly at the resonance energy for B.

    \item For an infinitesimally thin $\delta$-layer, $h(z)=\delta(z)$ and $H(Q)=1$, the maximum amplitude of the anomaly is $2r_{0}N_\text{2D}\left|\Delta fF_Q(\rho)\right|$ independent of the $Q$ sampled. For a Gaussian, $h(z)=1/\left(\sqrt{2\pi}\sigma\right)\exp(-[z/\sigma]^2/2)$, so that $H(Q)=\exp(-[Q\sigma]^2/2)$, implying that if $N_\text{2D}$ and $|F_Q(\rho)|$ are known by other means, the amplitude of the anomaly at any non-zero $Q$ will be reduced by the factor $H(Q)$, from which the $\delta$-layer thickness parameter $\delta=2\sqrt{2}\pi\sigma$ can be extracted. 

    \item Assuming that $F_Q(\rho)$ and $H(Q)$ vary slowly with~$Q$ on the scale of the oscillations from the $\delta$-layer, a moving average or low-pass filter with window of width $2\pi/d$ will remove the interference term in Eq.~(\ref{M_eq_Born_approx}), leaving an estimate $|F|^2_\mathrm{LF}$ of the strong zeroth-order term $|F_Q(\rho)|^2$. The interference term is then estimated as
    \begin{equation}
    \label{M_I(Q)_estimate}
    I(Q) \approx \frac{R(Q)}{r_F(Q)^2}-|F|^2_\mathrm{LF}.
    \end{equation}
    Thus we can approximate $\delta{\rho}$ as the inverse Fourier transform of the measured oscillations divided by $\sqrt{|F|^2_\mathrm{LF}} \approx |F_Q(\rho)|^2$, assuming that $F_Q(\rho)$ is imaginary for non-zero $Q$, which will be the case if $\rho(z)$ is a step function at $z=0$, a reasonable approximation for samples of the type considered here:
    \begin{equation}
    \bar{\delta{\rho}} = \mathrm{FT}^{-1} \left( \frac{I(Q)}{2\sqrt{|F|^2_\mathrm{LF}}} \right).
    \label{M_inverse_transform}
    \end{equation}
    Finally, for the Gaussian $\delta$-layer profile discussed under point 3., dividing Eq.~(\ref{M_eq_sym_interference}) by $\sqrt{|F|^2_\mathrm{LF}}$ gives:
    \begin{equation}
    \frac{I(Q)}{\sqrt{|F|^2_\mathrm{LF}}} = 
    2r_{0}N_\text{2D}|\Delta f|\exp\left(-\frac{(Q\sigma)^2}{2}\right)\cos(Qd - \phi),
    \label{M_eq_gauss_profile}
    \end{equation}
    where $\phi = \text{arg}(\Delta f)$.
\end{enumerate}

Instead of directly analyzing data as described above or via modelling software, it is also useful to consider differences $\Delta R(Q)=R(E',Q)-R(E,Q)$ for two different X-ray energies $E$ and $E'$ close to a type B atom resonance, such that changes in terms not containing B atom contributions can be omitted. In terms of the Born approximation this leads to:
\begin{equation}
\begin{aligned}
\frac{\Delta R(Q)}{r_F(Q)^2} = 2\Re\left[F_Q(\rho)\left(\bar{F_Q^{E'}}(\delta\tilde{\rho})-\bar{F_Q^E}(\delta\tilde{\rho})\right)\right],
\label{M_eq_Born_diff1}
\end{aligned}
\end{equation}
where $F_Q^{E'}(\rho)=F_Q^E(\rho)=F_Q(\rho)$ as it does not contain type B atoms.
This shows that measuring the difference in reflectometry at two energies near a $\delta$-layer's atomic absorption edge isolates the dopant's signal, dependent to first order only on the change in the dopant's atomic scattering factors at the absorption edge.

Applying the assumptions that led to Eq.~(\ref{M_eq_sym_interference}), we can recast Eq.~(\ref{M_eq_Born_diff1}) as 
\begin{equation}
\frac{\Delta R(Q)}{r_F(Q)^2} = 2r_{0}N_\text{2D} H(Q)\Re\left[ \left(\bar{\Delta f_{E'}}-\bar{\Delta f_{E}} \right)\exp(iQd)F_Q(\rho) \right].%,
\label{M_eq_Born_diff1_simplified}
\end{equation}
This matches Eq.~(\ref{M_eq_sym_interference}), with $I(Q) \to \Delta R(Q)/r_F(Q)^2$ and $\Delta f \to \Delta f_{E'}-\Delta f_{E}$, so Eq.~(\ref{M_inverse_transform})~\&~(\ref{M_eq_gauss_profile}) apply with these substitutions.
This resonant contrast analysis has the advantage of isolating the interference term~[Eq.~(\ref{M_eq_Born_diff1})] without needing to separate signals at different frequencies, as required when analysing reflectometry at a single energy [Eq.~(\ref{M_I(Q)_estimate})].

\noindent\textbf{Sample preparation.}
Four $2\times9$~mm$^2$ Si(001) samples were diced from Czochralski-grown wafers (bulk doping~$<5\times10^{14}$~cm$^{-3}$), and cleaned ultrasonically in acetone, followed by isopropyl alcohol. Each sample was thermally outgassed in vacuum (base pressure~\mbox{$<5\times10^{-10}$~mbar}) for~\textgreater~8~h at 600~$^{\circ}$C, then flash annealed multiple times at 1200~$^{\circ}$C using direct current resistive sample heating. This temperature was monitored using an infrared pyrometer (\mbox{IMPAC IGA50-LO plus}) with an uncertainty of $\pm30~^{\circ}$C.
Each sample was dosed with AsH$_3$, to the saturation As~density of $(1.6\pm0.3)\times10^{14}$~cm$^{-2}$ \cite{Stock_As_count}, and heated to 350~$^{\circ}$C for $1-2$~min to incorporate the arsenic into the silicon lattice \cite{Goh2005}. Afterwards, a $1-4$~nm silicon \textquote{locking layer} was deposited, without resistive sample heating, to confine the arsenic \cite{Keizer2015, Stock_As_count}. Three of the samples were then heated to~500~$^{\circ}$C for 15~s, to improve electrical activation, whereas sample \#1 was not heated. More silicon ($14-71$~nm depending on sample) was deposited, with the sample at 250~$^{\circ}$C. Silicon was deposited at a rate of $0.1-0.4$~nm/min using a silicon solid sublimation source (\mbox{SUSI-40}, MBE Komponenten GmbH). During deposition, the sample temperature was indirectly monitored by the sample resistance, while heating using a direct current resistive sample heater.
A control sample with a buried oxide layer was also grown. The process was as above, except without the flash anneal or AsH$_3$ dose, and the silicon deposition (15~nm) was done without sample heating.
All samples and parameters are shown in Tab.~ \ref{tab:XRR_samples}. The sample in Fig.~\ref{fig_resonance_contrast} and~\ref{fig_resonance} was etched with HF prior to the RCXR measurement to make the surface oxide layer thinner, which suppresses an additional slow oscillation to the reflectometry. 

\begin{table}[t]
    \centering
    \begin{tabular}{@{}m{0.3cm}m{1.1cm}m{1.1cm}m{0.9cm}m{1.3cm}m{1.3cm}m{1.8cm}@{}}
        \toprule
        \# & Dopant & Locking layer & 500~$^{\circ}$C anneal & $d_\text{SIMS}$ (nm) & $d_\text{XRR}$ (nm) & $\delta_\text{XRR}$ (nm) \\
        \midrule
        1 & As & Yes & No & 16.0$\pm$1.7 & 16.8$\pm$1.7 & 0.8~(0.6\text{--}1.2) \\
        2 & As & Yes & Yes & 15.3$\pm$1.7 & 17.9$\pm$1.7 & 0.6~(0.4\text{--}1.3) \\
        3 & As & Yes & Yes & 28.7$\pm$2.3 & 32$\pm$2 & 1.4~(1.0\text{--}2.0) \\
        4 & As & Yes & Yes & 70.5$\pm$3.0 & 77$\pm$3 & 1.2~(1.0\text{--}1.4) \\ 
        5 & SiO$_2$ & No & No & 13.4$\pm$1.8 & 15.3$\pm$0.8 & -- \\
        \bottomrule
    \end{tabular}
    \caption{Samples details, depths $d$ and thicknesses $\delta$ measured by SIMS and XRR. SIMS errors are from the uncertainty in the sputter rate. XRR depth errors are from the FWHM of the main peak in the signal's Fourier transform. The uncertainty on the thicknesses is the 95\% confidence interval of fits to Eq.~(\ref{eq_gauss_profile}). 
    }
    \label{tab:XRR_samples}
\end{table}

\begin{figure}
 \centering
 \includegraphics[width=\linewidth]{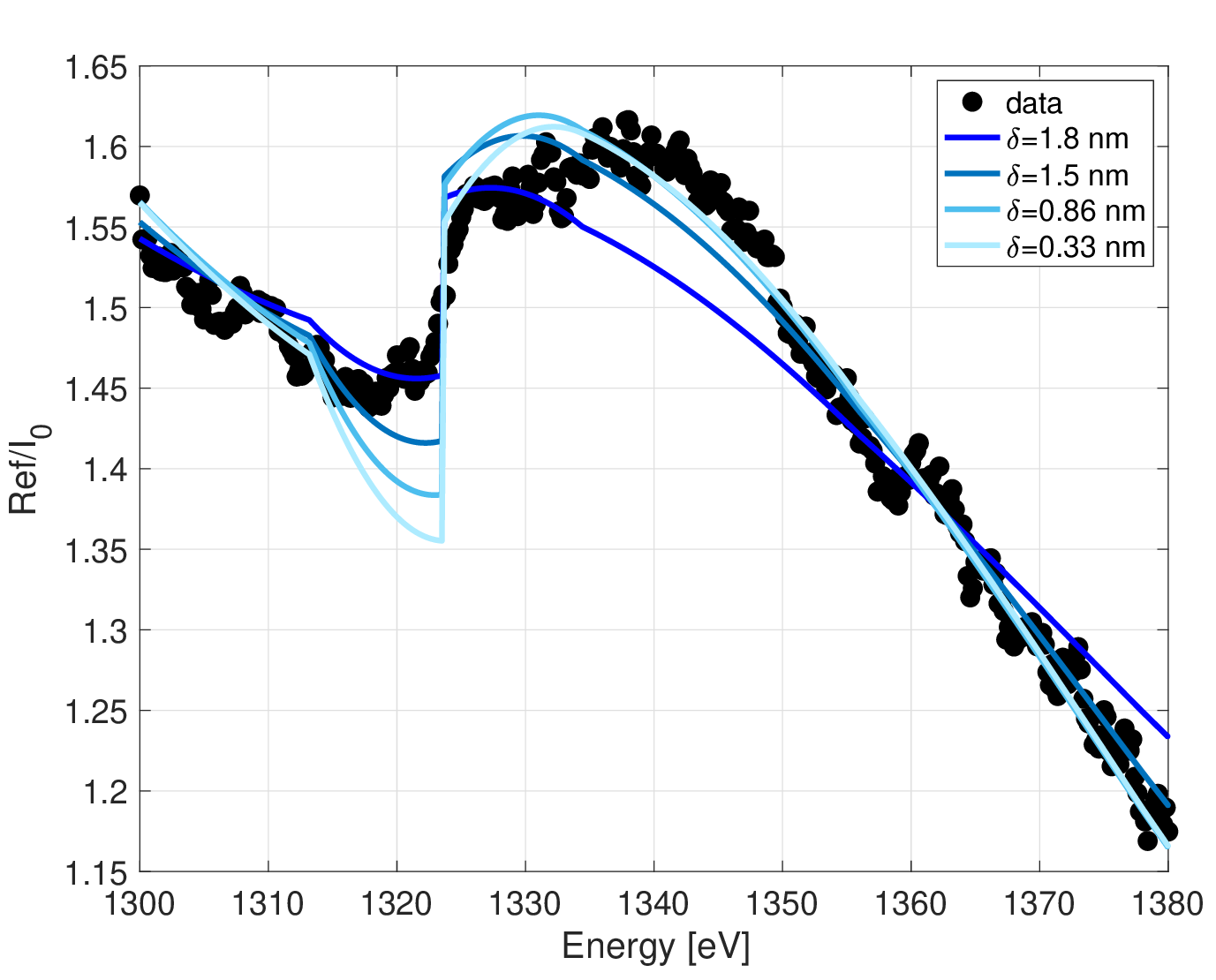}
\caption{\textbf{Arsenic $L_3$-edge fits.}
Black dots correspond to the reflectometry data shown in Fig.~\ref{fig_resonance_left}. The lines are fits to the data with colours corresponding to a Si:As layer thickness $\delta$ as indicated in the legend.}
\label{Spectrum_fits}
\end{figure}  

\noindent\textbf{X-ray scattering.}
RCXR was performed at the \mbox{RESOXS} endstation of the SIM beamline at the Swiss Light Source synchrotron of the Paul Scherrer Institute~\cite{doi:10.1063/1.3463200,Staub:fh5385}.
Samples were kept in high vacuum at~10$^{-8}$~mbar, and at room temperature. They were mounted on a rotatable holder such that the incidence angle $\theta$ could be swept from $0^{\circ}$ to $90^{\circ}$. The beam energy was set between 1200 and 1400~eV with linear polarization and a spot size set between 500 and 120~µm in the horizontal and 25 to 50~µm in the vertical direction.

\noindent\textbf{Arsenic $L_3$-edge reflectometry simulations.}
To assess the expected change in reflection $\Delta R/R$ at the arsenic $L_3$-edge, we use the DYNA program \cite{Dyna} to obtain the best fit to the resonant spectrum for different Si:As layer thicknesses. The resulting fits are shown in Fig.~\ref{Spectrum_fits}, each taking into account the oxide, silicon, and Si:As layer thicknesses, densities, and roughness. The silicon and oxide density is known, and the density of the Si:As layer is calculated from the 2D density $N_\text{2D}$ measured by X-ray fluorescence and the layer's thickness~$\delta$ ($N_\mathrm{3D}=N_\mathrm{2D}/\delta$), such that the simulated $\Delta R/R$ captures the larger arsenic density in thinner Si:As \mbox{$\delta$-layers}. The surface roughness obtained from the fits varies in the range $0.1-2~\text{\AA}$, in agreement with STM measurements where roughness was found to be in the range of 1~$\text{\AA}$.
The experimental values for $R(1320~\text{eV})$ and $R(1330~\text{eV})$ are taken as the mean of the data from 1312~to 1322~eV and 1330 to 1340~eV, respectively. The standard deviation is determined in the same intervals, yielding $\Delta R/R = 0.09\pm0.01$. $\Delta R$ of the simulation is taken to be $[\text{max}(R)-\text{min}(R)]$ in the interval~1320 to 1340~eV. The values for $\Delta R/R$ from the fits are shown as black dots on Fig.~\ref{fig_resonance_right} with a \textquote{shape preserving interpolant} line as a guide to the eye.

\begin{figure}[t]
    \centering
    \includegraphics[width=\linewidth]{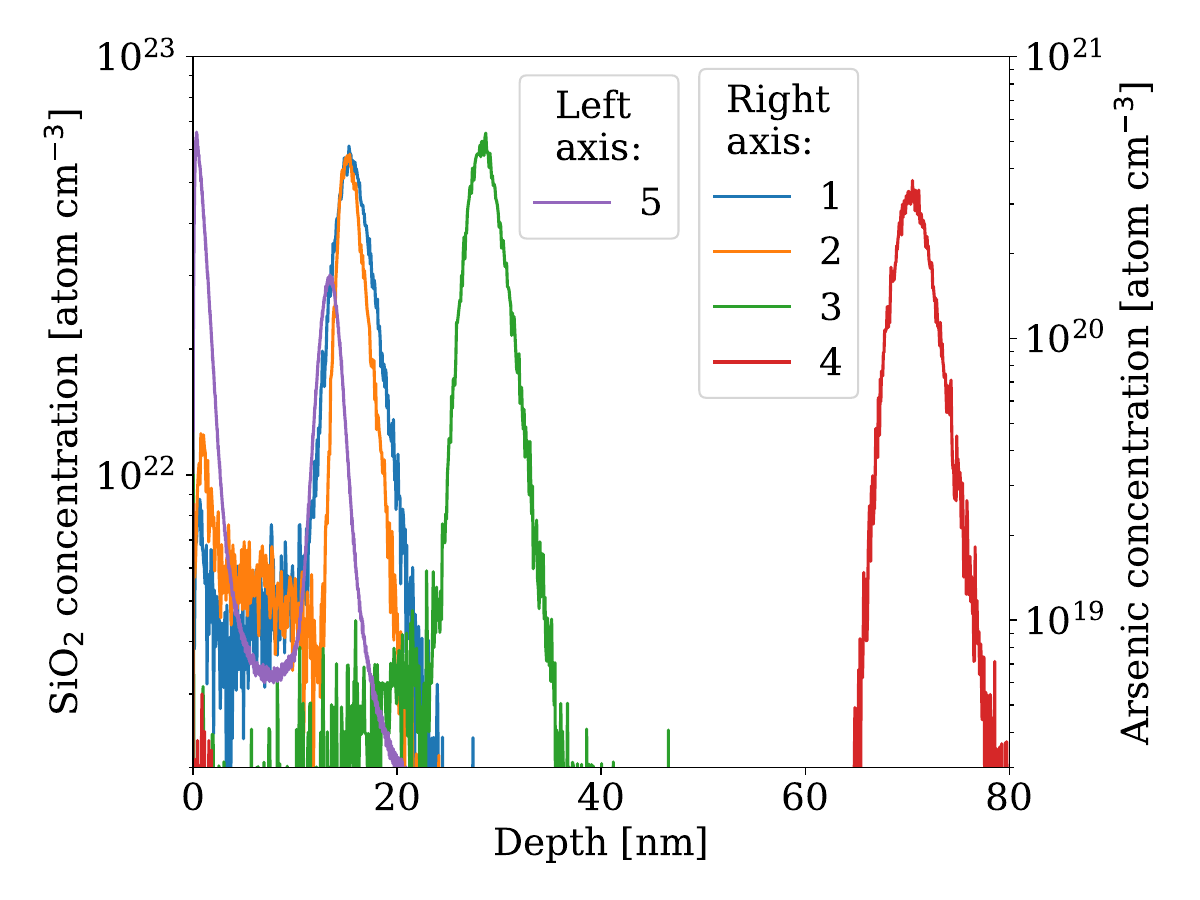}
    \caption{\textbf{SIMS depth profiles.} Arsenic depth profiles of the four Si:As $\delta$-layers of varying depth, along with the oxide profile of the buried oxide layer (see Tab.~\ref{tab:XRR_samples}).}
    \label{SIMS}
\end{figure}

\noindent\textbf{Secondary ion mass spectrometry.}
Time-of-flight SIMS (IONTOF ToF-SIMS5) was conducted on all samples with a~25~keV, 1~pA Bi$^+$ primary ion beam in high current bunch mode, and a 500~eV, 35--50~nA Cs$^+$ sputter beam. Depth profiles were made with a \mbox{300 $\times$ 300~Î¼m$^2$} sputter crater, within which the analytical region was the central~50 $\times$ 50~Î¼m$^2$ or 100 $\times$ 100~Î¼m$^2$. The sputter rate was determined by measuring the crater depth with an interference microscope (Zygo NewView NX2). The crater depth was measured from line profiles of the topography in different directions, where the uncertainty was estimated from the standard deviation of these measurements. The results are shown in Fig.~\ref{SIMS}.

\noindent\textbf{Scanning probe microscopy.}
STM was conducted using an Omicron variable temperature system. AFM was preformed with a Bruker Dimension Icon with a ScanAsyst-Air cantilever, using peak force tapping mode.\\

\FloatBarrier

\begin{acknowledgments}
We acknowledge the Paul Scherrer Institute, Villigen, Switzerland for provision of synchrotron radiation at the RESOXS endstation at the SIM beamline and the microXAS beamline of the Swiss Light Source. This project received funding from the European Research Council under the European Union's Horizon 2020 research and innovation program, within the Hidden, Entangled and Resonating Order (HERO) project with Grant Agreement 810451. The project was financially supported by the Engineering and Physical Sciences Research Council (EPSRC) Grant Numbers EP/M009564/1, EP/R034540/1, EP/V027700/1, and EP/W000520/1, as well as, Innovate UK Grant Number UKRI/75574. N.D. was partially supported by Swiss National Science Foundation Contract 175867.
E.S. received funding from the European Union's Horizon 2020 research and innovation program under the Marie Sklodowska-Curie Grant Agreement 884104 \mbox{(PSI-FELLOW-III-3i)}.
H.U. was supported by the National Centers of Competence in Research in Molecular Ultrafast Science and Technology (Grant Number~51NF40-183615) from the Swiss National Science Foundation, and the European Union's Horizon 2020 research and innovation program Marie Sklodowska-Curie Grant Agreement 801459 (FP-RESOMUS).
J.B, K.S., and P.C.C. were supported by the EPSRC Centre for Doctoral Training in Advanced Characterization of Materials (Grant Number~EP/L015277/1), and by the Paul Scherrer Institute.
\end{acknowledgments}

\bibliography{biblio}				

\begin{thebibliography}{60}%
\makeatletter
\providecommand \@ifxundefined [1]{%
 \@ifx{#1\undefined}
}%
\providecommand \@ifnum [1]{%
 \ifnum #1\expandafter \@firstoftwo
 \else \expandafter \@secondoftwo
 \fi
}%
\providecommand \@ifx [1]{%
 \ifx #1\expandafter \@firstoftwo
 \else \expandafter \@secondoftwo
 \fi
}%
\providecommand \natexlab [1]{#1}%
\providecommand \enquote  [1]{``#1''}%
\providecommand \bibnamefont  [1]{#1}%
\providecommand \bibfnamefont [1]{#1}%
\providecommand \citenamefont [1]{#1}%
\providecommand \href@noop [0]{\@secondoftwo}%
\providecommand \href [0]{\begingroup \@sanitize@url \@href}%
\providecommand \@href[1]{\@@startlink{#1}\@@href}%
\providecommand \@@href[1]{\endgroup#1\@@endlink}%
\providecommand \@sanitize@url [0]{\catcode `\\12\catcode `\$12\catcode `\&12\catcode `\#12\catcode `\^12\catcode `\_12\catcode `\%12\relax}%
\providecommand \@@startlink[1]{}%
\providecommand \@@endlink[0]{}%
\providecommand \url  [0]{\begingroup\@sanitize@url \@url }%
\providecommand \@url [1]{\endgroup\@href {#1}{\urlprefix }}%
\providecommand \urlprefix  [0]{URL }%
\providecommand \Eprint [0]{\href }%
\providecommand \doibase [0]{https://doi.org/}%
\providecommand \selectlanguage [0]{\@gobble}%
\providecommand \bibinfo  [0]{\@secondoftwo}%
\providecommand \bibfield  [0]{\@secondoftwo}%
\providecommand \translation [1]{[#1]}%
\providecommand \BibitemOpen [0]{}%
\providecommand \bibitemStop [0]{}%
\providecommand \bibitemNoStop [0]{.\EOS\space}%
\providecommand \EOS [0]{\spacefactor3000\relax}%
\providecommand \BibitemShut  [1]{\csname bibitem#1\endcsname}%
\let\auto@bib@innerbib\@empty
%</preamble>
\bibitem [{\citenamefont {Stroscio}\ and\ \citenamefont {Eigler}(1991)}]{STM_manip}%
  \BibitemOpen
  \bibfield  {author} {\bibinfo {author} {\bibfnamefont {J.~A.}\ \bibnamefont {Stroscio}}\ and\ \bibinfo {author} {\bibfnamefont {D.~M.}\ \bibnamefont {Eigler}},\ }\bibfield  {title} {\bibinfo {title} {Atomic and molecular manipulation with the scanning tunneling microscope},\ }\href {https://doi.org/10.1126/science.254.5036.1319} {\bibfield  {journal} {\bibinfo  {journal} {Science}\ }\textbf {\bibinfo {volume} {254}},\ \bibinfo {pages} {1319} (\bibinfo {year} {1991})}\BibitemShut {NoStop}%
\bibitem [{\citenamefont {Crommie}\ \emph {et~al.}(1993)\citenamefont {Crommie}, \citenamefont {Lutz},\ and\ \citenamefont {Eigler}}]{STM_corrals}%
  \BibitemOpen
  \bibfield  {author} {\bibinfo {author} {\bibfnamefont {M.~F.}\ \bibnamefont {Crommie}}, \bibinfo {author} {\bibfnamefont {C.~P.}\ \bibnamefont {Lutz}},\ and\ \bibinfo {author} {\bibfnamefont {D.~M.}\ \bibnamefont {Eigler}},\ }\bibfield  {title} {\bibinfo {title} {Confinement of electrons to quantum corrals on a metal surface},\ }\href {https://doi.org/10.1126/science.262.5131.218} {\bibfield  {journal} {\bibinfo  {journal} {Science}\ }\textbf {\bibinfo {volume} {262}},\ \bibinfo {pages} {218} (\bibinfo {year} {1993})}\BibitemShut {NoStop}%
\bibitem [{\citenamefont {Ruess}\ \emph {et~al.}(2004)\citenamefont {Ruess} \emph {et~al.}}]{STM_P}%
  \BibitemOpen
  \bibfield  {author} {\bibinfo {author} {\bibfnamefont {F.~J.}\ \bibnamefont {Ruess}} \emph {et~al.},\ }\bibfield  {title} {\bibinfo {title} {Toward atomic-scale device fabrication in silicon using scanning probe microscopy},\ }\href {https://doi.org/10.1021/nl048808v} {\bibfield  {journal} {\bibinfo  {journal} {Nano Lett.}\ }\textbf {\bibinfo {volume} {4}},\ \bibinfo {pages} {1969} (\bibinfo {year} {2004})}\BibitemShut {NoStop}%
\bibitem [{\citenamefont {Fuechsle}\ \emph {et~al.}(2012)\citenamefont {Fuechsle} \emph {et~al.}}]{Single_atom_transistor}%
  \BibitemOpen
  \bibfield  {author} {\bibinfo {author} {\bibfnamefont {M.}~\bibnamefont {Fuechsle}} \emph {et~al.},\ }\bibfield  {title} {\bibinfo {title} {Single-atom transistor},\ }\href {https://doi.org/10.1038/nnano.2012.21} {\bibfield  {journal} {\bibinfo  {journal} {Nat. Nanotechnol.}\ }\textbf {\bibinfo {volume} {7}},\ \bibinfo {pages} {242} (\bibinfo {year} {2012})}\BibitemShut {NoStop}%
\bibitem [{\citenamefont {Stock}\ \emph {et~al.}(2020)\citenamefont {Stock} \emph {et~al.}}]{Stock_As_count}%
  \BibitemOpen
  \bibfield  {author} {\bibinfo {author} {\bibfnamefont {T.~J.~Z.}\ \bibnamefont {Stock}} \emph {et~al.},\ }\bibfield  {title} {\bibinfo {title} {Atomic-scale patterning of arsenic in silicon by scanning tunneling microscopy},\ }\href {https://doi.org/10.1021/acsnano.9b08943} {\bibfield  {journal} {\bibinfo  {journal} {ACS Nano}\ }\textbf {\bibinfo {volume} {14}},\ \bibinfo {pages} {3316} (\bibinfo {year} {2020})}\BibitemShut {NoStop}%
\bibitem [{\citenamefont {Dwyer}\ \emph {et~al.}(2021)\citenamefont {Dwyer} \emph {et~al.}}]{Dwyer:2021aa}%
  \BibitemOpen
  \bibfield  {author} {\bibinfo {author} {\bibfnamefont {K.~J.}\ \bibnamefont {Dwyer}} \emph {et~al.},\ }\bibfield  {title} {\bibinfo {title} {B-doped $\delta$-layers and nanowires from area-selective deposition of {$\rm BCl_3$} on {$\rm Si(100)$}},\ }\href {https://doi.org/10.1021/acsami.1c10616} {\bibfield  {journal} {\bibinfo  {journal} {ACS Appl. Mater. Interfaces}\ }\textbf {\bibinfo {volume} {13}},\ \bibinfo {pages} {41275} (\bibinfo {year} {2021})}\BibitemShut {NoStop}%
\bibitem [{\citenamefont {Jeong}\ \emph {et~al.}(2018)\citenamefont {Jeong} \emph {et~al.}}]{Samsung_FET}%
  \BibitemOpen
  \bibfield  {author} {\bibinfo {author} {\bibfnamefont {W.~C.}\ \bibnamefont {Jeong}} \emph {et~al.},\ }\bibfield  {title} {\bibinfo {title} {True 7~nm {P}latform {T}echnology featuring {S}mallest {F}in{FET} and {S}mallest {SRAM} cell by {EUV}, {S}pecial {C}onstructs and 3rd {G}eneration {S}ingle {D}iffusion {B}reak},\ }in\ \href {https://doi.org/10.1109/VLSIT.2018.8510682} {\emph {\bibinfo {booktitle} {2018 IEEE Symposium on VLSI Technology}}}\ (\bibinfo {year} {2018})\ pp.\ \bibinfo {pages} {59--60}\BibitemShut {NoStop}%
\bibitem [{\citenamefont {Wang}\ \emph {et~al.}(2018)\citenamefont {Wang} \emph {et~al.}}]{Wang2018}%
  \BibitemOpen
  \bibfield  {author} {\bibinfo {author} {\bibfnamefont {X.}~\bibnamefont {Wang}} \emph {et~al.},\ }\bibfield  {title} {\bibinfo {title} {Quantifying atom-scale dopant movement and electrical activation in {Si:P} monolayers},\ }\href {https://doi.org/10.1039/c7nr07777g} {\bibfield  {journal} {\bibinfo  {journal} {Nanoscale}\ }\textbf {\bibinfo {volume} {10}},\ \bibinfo {pages} {4488} (\bibinfo {year} {2018})}\BibitemShut {NoStop}%
\bibitem [{\citenamefont {Chang}\ and\ \citenamefont {Lauhon}(2018)}]{atome_probe}%
  \BibitemOpen
  \bibfield  {author} {\bibinfo {author} {\bibfnamefont {A.~S.}\ \bibnamefont {Chang}}\ and\ \bibinfo {author} {\bibfnamefont {L.~J.}\ \bibnamefont {Lauhon}},\ }\bibfield  {title} {\bibinfo {title} {Atom probe tomography of nanoscale architectures in functional materials for electronic and photonic applications},\ }\href {https://doi.org/https://doi.org/10.1016/j.cossms.2018.09.002} {\bibfield  {journal} {\bibinfo  {journal} {Curr. Opin. Solid State Mater. Sci.}\ }\textbf {\bibinfo {volume} {22}},\ \bibinfo {pages} {171} (\bibinfo {year} {2018})}\BibitemShut {NoStop}%
\bibitem [{\citenamefont {Werner}\ and\ \citenamefont {Boudewijn}(1984)}]{SIMS}%
  \BibitemOpen
  \bibfield  {author} {\bibinfo {author} {\bibfnamefont {H.~W.}\ \bibnamefont {Werner}}\ and\ \bibinfo {author} {\bibfnamefont {P.~R.}\ \bibnamefont {Boudewijn}},\ }\bibfield  {title} {\bibinfo {title} {A comparison of {SIMS} with other techniques based on ion-beam solid interactions},\ }\href {https://doi.org/https://doi.org/10.1016/0042-207X(84)90111-8} {\bibfield  {journal} {\bibinfo  {journal} {Vacuum}\ }\textbf {\bibinfo {volume} {34}},\ \bibinfo {pages} {83} (\bibinfo {year} {1984})}\BibitemShut {NoStop}%
\bibitem [{\citenamefont {D'Anna}\ \emph {et~al.}(2023)\citenamefont {D'Anna} \emph {et~al.}}]{DAnna_Xray_fluorescence}%
  \BibitemOpen
  \bibfield  {author} {\bibinfo {author} {\bibfnamefont {N.}~\bibnamefont {D'Anna}} \emph {et~al.},\ }\bibfield  {title} {\bibinfo {title} {Non-destructive {X}-ray imaging of patterned delta-layer devices in silicon},\ }\href {https://doi.org/https://doi.org/10.1002/aelm.202201212} {\bibfield  {journal} {\bibinfo  {journal} {Adv. Electron. Mater.}\ }\textbf {\bibinfo {volume} {9}},\ \bibinfo {pages} {2201212} (\bibinfo {year} {2023})}\BibitemShut {NoStop}%
\bibitem [{\citenamefont {Masteghin}\ \emph {et~al.}(2024)\citenamefont {Masteghin} \emph {et~al.}}]{https://doi.org/10.1002/smtd.202301610}%
  \BibitemOpen
  \bibfield  {author} {\bibinfo {author} {\bibfnamefont {M.~G.}\ \bibnamefont {Masteghin}} \emph {et~al.},\ }\bibfield  {title} {\bibinfo {title} {{Benchmarking of X-Ray fluorescence microscopy with ion beam implanted samples showing detection sensitivity of hundreds of atoms}},\ }\href {https://doi.org/https://doi.org/10.1002/smtd.202301610} {\bibfield  {journal} {\bibinfo  {journal} {Small Methods}\ }\textbf {\bibinfo {volume} {2024}},\ \bibinfo {pages} {2301610} (\bibinfo {year} {2024})}\BibitemShut {NoStop}%
\bibitem [{\citenamefont {Sch{\"u}lli}\ and\ \citenamefont {Leake}(2018)}]{SCHULLI2018188}%
  \BibitemOpen
  \bibfield  {author} {\bibinfo {author} {\bibfnamefont {T.~U.}\ \bibnamefont {Sch{\"u}lli}}\ and\ \bibinfo {author} {\bibfnamefont {S.~J.}\ \bibnamefont {Leake}},\ }\bibfield  {title} {\bibinfo {title} {X-ray nanobeam diffraction imaging of materials},\ }\href {https://doi.org/https://doi.org/10.1016/j.cossms.2018.09.003} {\bibfield  {journal} {\bibinfo  {journal} {Curr. Opin. Solid State Mater. Sci.}\ }\textbf {\bibinfo {volume} {22}},\ \bibinfo {pages} {188} (\bibinfo {year} {2018})}\BibitemShut {NoStop}%
\bibitem [{\citenamefont {Constantinou}\ \emph {et~al.}(2023)\citenamefont {Constantinou} \emph {et~al.}}]{Procopi}%
  \BibitemOpen
  \bibfield  {author} {\bibinfo {author} {\bibfnamefont {P.}~\bibnamefont {Constantinou}} \emph {et~al.},\ }\bibfield  {title} {\bibinfo {title} {Momentum-space imaging of ultra-thin electron liquids in \mbox{$\delta$-doped} silicon},\ }\href {https://doi.org/https://doi.org/10.1002/advs.202302101} {\bibfield  {journal} {\bibinfo  {journal} {Adv. Sci.}\ }\textbf {\bibinfo {volume} {10}},\ \bibinfo {pages} {2302101} (\bibinfo {year} {2023})}\BibitemShut {NoStop}%
\bibitem [{\citenamefont {Katzenmeyer}\ \emph {et~al.}(2020)\citenamefont {Katzenmeyer} \emph {et~al.}}]{Katzenmeyer2020}%
  \BibitemOpen
  \bibfield  {author} {\bibinfo {author} {\bibfnamefont {A.~M.}\ \bibnamefont {Katzenmeyer}} \emph {et~al.},\ }\bibfield  {title} {\bibinfo {title} {Assessing atomically thin delta-doping of silicon using mid-infrared ellipsometry},\ }\href {https://doi.org/10.1557/jmr.2020.155} {\bibfield  {journal} {\bibinfo  {journal} {J. Mater. Res.}\ }\textbf {\bibinfo {volume} {35}},\ \bibinfo {pages} {2098} (\bibinfo {year} {2020})}\BibitemShut {NoStop}%
\bibitem [{\citenamefont {Young}\ \emph {et~al.}(2023)\citenamefont {Young} \emph {et~al.}}]{Young2023}%
  \BibitemOpen
  \bibfield  {author} {\bibinfo {author} {\bibfnamefont {S.~M.}\ \bibnamefont {Young}} \emph {et~al.},\ }\bibfield  {title} {\bibinfo {title} {{Suppression of midinfrared plasma resonance due to quantum confinement in $\delta$-doped silicon}},\ }\href {https://doi.org/10.1103/PHYSREVAPPLIED.20.024043/FIGURES/6/MEDIUM} {\bibfield  {journal} {\bibinfo  {journal} {Phys. Rev. Appl.}\ }\textbf {\bibinfo {volume} {20}},\ \bibinfo {pages} {024043} (\bibinfo {year} {2023})}\BibitemShut {NoStop}%
\bibitem [{\citenamefont {Barber}\ \emph {et~al.}(2022)\citenamefont {Barber}, \citenamefont {Ma},\ and\ \citenamefont {Shen}}]{MIM}%
  \BibitemOpen
  \bibfield  {author} {\bibinfo {author} {\bibfnamefont {M.~E.}\ \bibnamefont {Barber}}, \bibinfo {author} {\bibfnamefont {E.~Y.}\ \bibnamefont {Ma}},\ and\ \bibinfo {author} {\bibfnamefont {Z.-X.}\ \bibnamefont {Shen}},\ }\bibfield  {title} {\bibinfo {title} {Microwave impedance microscopy and its application to quantum materials},\ }\href {https://doi.org/10.1038/s42254-021-00386-3} {\bibfield  {journal} {\bibinfo  {journal} {Nat. Rev. Phys.}\ }\textbf {\bibinfo {volume} {4}},\ \bibinfo {pages} {61} (\bibinfo {year} {2022})}\BibitemShut {NoStop}%
\bibitem [{\citenamefont {Gramse}\ \emph {et~al.}(2017)\citenamefont {Gramse} \emph {et~al.}}]{NondestructiveAtomThin}%
  \BibitemOpen
  \bibfield  {author} {\bibinfo {author} {\bibfnamefont {G.}~\bibnamefont {Gramse}} \emph {et~al.},\ }\bibfield  {title} {\bibinfo {title} {Nondestructive imaging of atomically thin nanostructures buried in silicon},\ }\href {https://doi.org/10.1126/sciadv.1602586} {\bibfield  {journal} {\bibinfo  {journal} {Sci. Adv.}\ }\textbf {\bibinfo {volume} {3}},\ \bibinfo {pages} {e1602586} (\bibinfo {year} {2017})}\BibitemShut {NoStop}%
\bibitem [{\citenamefont {Gramse}\ \emph {et~al.}(2020)\citenamefont {Gramse} \emph {et~al.}}]{bb_EFM}%
  \BibitemOpen
  \bibfield  {author} {\bibinfo {author} {\bibfnamefont {G.}~\bibnamefont {Gramse}} \emph {et~al.},\ }\bibfield  {title} {\bibinfo {title} {Nanoscale imaging of mobile carriers and trapped charges in delta doped silicon p--n junctions},\ }\href {https://doi.org/10.1038/s41928-020-0450-8} {\bibfield  {journal} {\bibinfo  {journal} {Nat. Electron.}\ }\textbf {\bibinfo {volume} {3}},\ \bibinfo {pages} {531} (\bibinfo {year} {2020})}\BibitemShut {NoStop}%
\bibitem [{\citenamefont {Ng}\ \emph {et~al.}(2020)\citenamefont {Ng} \emph {et~al.}}]{doi:10.1021/acsnano.0c00736}%
  \BibitemOpen
  \bibfield  {author} {\bibinfo {author} {\bibfnamefont {K.~S.~H.}\ \bibnamefont {Ng}} \emph {et~al.},\ }\bibfield  {title} {\bibinfo {title} {Scanned single-electron probe inside a silicon electronic device},\ }\href {https://doi.org/10.1021/acsnano.0c00736} {\bibfield  {journal} {\bibinfo  {journal} {ACS Nano}\ }\textbf {\bibinfo {volume} {14}},\ \bibinfo {pages} {9449} (\bibinfo {year} {2020})}\BibitemShut {NoStop}%
\bibitem [{\citenamefont {Amsel}\ and\ \citenamefont {Lanford}(1984)}]{osti_5977235}%
  \BibitemOpen
  \bibfield  {author} {\bibinfo {author} {\bibfnamefont {G.}~\bibnamefont {Amsel}}\ and\ \bibinfo {author} {\bibfnamefont {W.~A.}\ \bibnamefont {Lanford}},\ }\bibfield  {title} {\bibinfo {title} {Nuclear reaction techniques in materials analysis},\ }\href {https://www.osti.gov/biblio/5977235} {\bibfield  {journal} {\bibinfo  {journal} {Annu. Rev. Nucl. Part. Sci.}\ }\textbf {\bibinfo {volume} {34}} (\bibinfo {year} {1984})}\BibitemShut {NoStop}%
\bibitem [{\citenamefont {Goncharova}(2020)}]{GONCHAROVA2020441}%
  \BibitemOpen
  \bibfield  {author} {\bibinfo {author} {\bibfnamefont {L.~V.}\ \bibnamefont {Goncharova}},\ }\bibfield  {title} {\bibinfo {title} {{Rutherford backscattering spectroscopy (RBS) and medium energy ion scattering (MEIS)}},\ }in\ \href {https://doi.org/https://doi.org/10.1016/B978-0-12-814182-3.00023-7} {\emph {\bibinfo {booktitle} {Characterization of Nanoparticles}}}\ (\bibinfo  {publisher} {Elsevier},\ \bibinfo {year} {2020})\ pp.\ \bibinfo {pages} {441--455}\BibitemShut {NoStop}%
\bibitem [{\citenamefont {Trombini}\ \emph {et~al.}(2019)\citenamefont {Trombini} \emph {et~al.}}]{TROMBINI2019137536}%
  \BibitemOpen
  \bibfield  {author} {\bibinfo {author} {\bibfnamefont {H.}~\bibnamefont {Trombini}} \emph {et~al.},\ }\bibfield  {title} {\bibinfo {title} {{Profiling As plasma doped Si/SiO$_2$ with molecular ions}},\ }\href {https://doi.org/https://doi.org/10.1016/j.tsf.2019.137536} {\bibfield  {journal} {\bibinfo  {journal} {Thin Solid Films}\ }\textbf {\bibinfo {volume} {692}},\ \bibinfo {pages} {137536} (\bibinfo {year} {2019})}\BibitemShut {NoStop}%
\bibitem [{\citenamefont {Sanchez}\ \emph {et~al.}(2011)\citenamefont {Sanchez} \emph {et~al.}}]{SANCHEZ2011654}%
  \BibitemOpen
  \bibfield  {author} {\bibinfo {author} {\bibfnamefont {D.}~\bibnamefont {Sanchez}} \emph {et~al.},\ }\bibfield  {title} {\bibinfo {title} {{Structural characterization of Pb nanoislands in SiO$_2$/Si interface synthesized by ion implantation through MEIS analysis}},\ }\href {https://doi.org/https://doi.org/10.1016/j.susc.2010.12.011} {\bibfield  {journal} {\bibinfo  {journal} {Surf. Sci.}\ }\textbf {\bibinfo {volume} {605}},\ \bibinfo {pages} {654} (\bibinfo {year} {2011})}\BibitemShut {NoStop}%
\bibitem [{\citenamefont {Holler}\ \emph {et~al.}(2017)\citenamefont {Holler} \emph {et~al.}}]{Holler:2017aa}%
  \BibitemOpen
  \bibfield  {author} {\bibinfo {author} {\bibfnamefont {M.}~\bibnamefont {Holler}} \emph {et~al.},\ }\bibfield  {title} {\bibinfo {title} {High-resolution non-destructive three-dimensional imaging of integrated circuits},\ }\href {https://doi.org/10.1038/nature21698} {\bibfield  {journal} {\bibinfo  {journal} {Nature}\ }\textbf {\bibinfo {volume} {543}},\ \bibinfo {pages} {402} (\bibinfo {year} {2017})}\BibitemShut {NoStop}%
\bibitem [{\citenamefont {de~Jonge}\ \emph {et~al.}(2010)\citenamefont {de~Jonge} \emph {et~al.}}]{3D_tomo}%
  \BibitemOpen
  \bibfield  {author} {\bibinfo {author} {\bibfnamefont {M.~D.}\ \bibnamefont {de~Jonge}} \emph {et~al.},\ }\bibfield  {title} {\bibinfo {title} {Quantitative 3{D} elemental microtomography of \textit{{C}yclotella meneghiniana} at 400-nm resolution},\ }\href {https://doi.org/10.1073/pnas.1001469107} {\bibfield  {journal} {\bibinfo  {journal} {Proc. Natl. Acad. Sci. U.S.A.}\ }\textbf {\bibinfo {volume} {107}},\ \bibinfo {pages} {15676} (\bibinfo {year} {2010})}\BibitemShut {NoStop}%
\bibitem [{\citenamefont {Zhou}\ and\ \citenamefont {Chen}(1995)}]{ZHOU1995223}%
  \BibitemOpen
  \bibfield  {author} {\bibinfo {author} {\bibfnamefont {X.-L.}\ \bibnamefont {Zhou}}\ and\ \bibinfo {author} {\bibfnamefont {S.-H.}\ \bibnamefont {Chen}},\ }\bibfield  {title} {\bibinfo {title} {Theoretical foundation of {X}-ray and neutron reflectometry},\ }\href {https://doi.org/https://doi.org/10.1016/0370-1573(94)00110-O} {\bibfield  {journal} {\bibinfo  {journal} {Phys. Rep.}\ }\textbf {\bibinfo {volume} {257}},\ \bibinfo {pages} {223} (\bibinfo {year} {1995})}\BibitemShut {NoStop}%
\bibitem [{\citenamefont {Slijkerman}\ \emph {et~al.}(1990)\citenamefont {Slijkerman} \emph {et~al.}}]{10.1063/1.347047}%
  \BibitemOpen
  \bibfield  {author} {\bibinfo {author} {\bibfnamefont {W.~F.~J.}\ \bibnamefont {Slijkerman}} \emph {et~al.},\ }\bibfield  {title} {\bibinfo {title} {{X‐ray reflectivity of an Sb delta‐doping layer in silicon}},\ }\href {https://doi.org/10.1063/1.347047} {\bibfield  {journal} {\bibinfo  {journal} {J. Appl. Phys.}\ }\textbf {\bibinfo {volume} {68}},\ \bibinfo {pages} {5105} (\bibinfo {year} {1990})}\BibitemShut {NoStop}%
\bibitem [{\citenamefont {Macke}\ \emph {et~al.}(2014)\citenamefont {Macke} \emph {et~al.}}]{Macke2014}%
  \BibitemOpen
  \bibfield  {author} {\bibinfo {author} {\bibfnamefont {S.}~\bibnamefont {Macke}} \emph {et~al.},\ }\bibfield  {title} {\bibinfo {title} {Element specific monolayer depth profiling},\ }\href {https://doi.org/10.1002/ADMA.201402028} {\bibfield  {journal} {\bibinfo  {journal} {Adv. Mater.}\ }\textbf {\bibinfo {volume} {26}},\ \bibinfo {pages} {6554} (\bibinfo {year} {2014})}\BibitemShut {NoStop}%
\bibitem [{\citenamefont {Parratt}(1954)}]{Parratt}%
  \BibitemOpen
  \bibfield  {author} {\bibinfo {author} {\bibfnamefont {L.~G.}\ \bibnamefont {Parratt}},\ }\bibfield  {title} {\bibinfo {title} {Surface studies of solids by total reflection of {X}-rays},\ }\href {https://doi.org/10.1103/PhysRev.95.359} {\bibfield  {journal} {\bibinfo  {journal} {Phys. Rev.}\ }\textbf {\bibinfo {volume} {95}},\ \bibinfo {pages} {359} (\bibinfo {year} {1954})}\BibitemShut {NoStop}%
\bibitem [{\citenamefont {Tolan}\ and\ \citenamefont {Press}(1998)}]{Xray_reflect_review}%
  \BibitemOpen
  \bibfield  {author} {\bibinfo {author} {\bibfnamefont {M.}~\bibnamefont {Tolan}}\ and\ \bibinfo {author} {\bibfnamefont {W.}~\bibnamefont {Press}},\ }\bibfield  {title} {\bibinfo {title} {X-ray and neutron reflectivity},\ }\href {https://doi.org/doi:10.1524/zkri.1998.213.6.319} {\bibfield  {journal} {\bibinfo  {journal} {Z. Kristallogr. - Cryst. Mater.}\ }\textbf {\bibinfo {volume} {213}},\ \bibinfo {pages} {319} (\bibinfo {year} {1998})}\BibitemShut {NoStop}%
\bibitem [{\citenamefont {Caticha}(1995)}]{Caticha1995}%
  \BibitemOpen
  \bibfield  {author} {\bibinfo {author} {\bibfnamefont {A.}~\bibnamefont {Caticha}},\ }\bibfield  {title} {\bibinfo {title} {{Reflection and transmission of X-rays by graded interfaces}},\ }\href {https://doi.org/10.1103/PhysRevB.52.9214} {\bibfield  {journal} {\bibinfo  {journal} {Phys. Rev. B}\ }\textbf {\bibinfo {volume} {52}},\ \bibinfo {pages} {9214} (\bibinfo {year} {1995})}\BibitemShut {NoStop}%
\bibitem [{\citenamefont {Als-Nielsen}\ and\ \citenamefont {McMorrow}(2011)}]{AlsNielsen_McMorrow}%
  \BibitemOpen
  \bibfield  {author} {\bibinfo {author} {\bibfnamefont {J.}~\bibnamefont {Als-Nielsen}}\ and\ \bibinfo {author} {\bibfnamefont {D.}~\bibnamefont {McMorrow}},\ }\bibinfo {title} {Refraction and reflection from interfaces},\ in\ \href {https://doi.org/https://doi.org/10.1002/9781119998365.ch5} {\emph {\bibinfo {booktitle} {Elements of Modern X‐ray Physics}}}\ (\bibinfo  {publisher} {Wiley},\ \bibinfo {year} {2011})\ pp.\ \bibinfo {pages} {147--205}\BibitemShut {NoStop}%
\bibitem [{\citenamefont {Al-Bayati}\ \emph {et~al.}(1991)\citenamefont {Al-Bayati}, \citenamefont {Orrman-Rossiter}, \citenamefont {van~den Berg},\ and\ \citenamefont {Armour}}]{Al-Bayati1991}%
  \BibitemOpen
  \bibfield  {author} {\bibinfo {author} {\bibfnamefont {A.~H.}\ \bibnamefont {Al-Bayati}}, \bibinfo {author} {\bibfnamefont {K.~G.}\ \bibnamefont {Orrman-Rossiter}}, \bibinfo {author} {\bibfnamefont {J.~A.}\ \bibnamefont {van~den Berg}},\ and\ \bibinfo {author} {\bibfnamefont {D.~G.}\ \bibnamefont {Armour}},\ }\bibfield  {title} {\bibinfo {title} {{Composition and structure of the native Si oxide by high depth resolution medium energy ion scatering}},\ }\href {https://doi.org/10.1016/0039-6028(91)90214-D} {\bibfield  {journal} {\bibinfo  {journal} {Surf. Sci.}\ }\textbf {\bibinfo {volume} {241}},\ \bibinfo {pages} {91} (\bibinfo {year} {1991})}\BibitemShut {NoStop}%
\bibitem [{\citenamefont {Henke}\ \emph {et~al.}(1993)\citenamefont {Henke}, \citenamefont {Gullikson},\ and\ \citenamefont {Davis}}]{HENKE1993181}%
  \BibitemOpen
  \bibfield  {author} {\bibinfo {author} {\bibfnamefont {B.~L.}\ \bibnamefont {Henke}}, \bibinfo {author} {\bibfnamefont {E.~M.}\ \bibnamefont {Gullikson}},\ and\ \bibinfo {author} {\bibfnamefont {J.~C.}\ \bibnamefont {Davis}},\ }\bibfield  {title} {\bibinfo {title} {X-ray interactions: Photoabsorption, scattering, transmission, and reflection at ${E} = 50-30,000$ e{V}, ${Z} = 1-92$},\ }\href {https://doi.org/https://doi.org/10.1006/adnd.1993.1013} {\bibfield  {journal} {\bibinfo  {journal} {At. Data Nucl. Data Tables}\ }\textbf {\bibinfo {volume} {54}},\ \bibinfo {pages} {181} (\bibinfo {year} {1993})}\BibitemShut {NoStop}%
\bibitem [{\citenamefont {Nevot}\ and\ \citenamefont {Croce}(1980)}]{nevot1980characterization}%
  \BibitemOpen
  \bibfield  {author} {\bibinfo {author} {\bibfnamefont {L.}~\bibnamefont {Nevot}}\ and\ \bibinfo {author} {\bibfnamefont {P.}~\bibnamefont {Croce}},\ }\bibfield  {title} {\bibinfo {title} {Characterization of surfaces by grazing {X}-ray reflection--application to the study of polishing of some silicate glasses},\ }\href@noop {} {\bibfield  {journal} {\bibinfo  {journal} {Rev. Phys. Appl.}\ }\textbf {\bibinfo {volume} {15}},\ \bibinfo {pages} {761} (\bibinfo {year} {1980})}\BibitemShut {NoStop}%
\bibitem [{\citenamefont {Goh}\ and\ \citenamefont {Simmons}(2009)}]{Goh2009}%
  \BibitemOpen
  \bibfield  {author} {\bibinfo {author} {\bibfnamefont {K.~E.~J.}\ \bibnamefont {Goh}}\ and\ \bibinfo {author} {\bibfnamefont {M.~Y.}\ \bibnamefont {Simmons}},\ }\bibfield  {title} {\bibinfo {title} {Impact of {Si} growth rate on coherent electron transport in {Si:P} delta-doped devices},\ }\href {https://doi.org/10.1063/1.3245313} {\bibfield  {journal} {\bibinfo  {journal} {Appl. Phys. Lett.}\ }\textbf {\bibinfo {volume} {95}},\ \bibinfo {pages} {142104} (\bibinfo {year} {2009})}\BibitemShut {NoStop}%
\bibitem [{\citenamefont {Chantler}\ \emph {et~al.}(2005)\citenamefont {Chantler} \emph {et~al.}}]{NIST_table}%
  \BibitemOpen
  \bibfield  {author} {\bibinfo {author} {\bibfnamefont {C.~T.}\ \bibnamefont {Chantler}} \emph {et~al.},\ }\bibfield  {title} {\bibinfo {title} {{X}-ray form factor, attenuation and scattering tables (version 2.1)},\ }\href {https://doi.org/10.18434/T4HS32} {10.18434/T4HS32} (\bibinfo {year} {2005})\BibitemShut {NoStop}%
\bibitem [{\citenamefont {Elzo}\ \emph {et~al.}(2012)\citenamefont {Elzo} \emph {et~al.}}]{Dyna}%
  \BibitemOpen
  \bibfield  {author} {\bibinfo {author} {\bibfnamefont {M.}~\bibnamefont {Elzo}} \emph {et~al.},\ }\bibfield  {title} {\bibinfo {title} {X-ray resonant magnetic reflectivity of stratified magnetic structures: {E}igenwave formalism and application to a {W/Fe/W} trilayer},\ }\href {https://doi.org/https://doi.org/10.1016/j.jmmm.2011.07.019} {\bibfield  {journal} {\bibinfo  {journal} {J. Magn. Magn. Mater.}\ }\textbf {\bibinfo {volume} {324}},\ \bibinfo {pages} {105} (\bibinfo {year} {2012})}\BibitemShut {NoStop}%
\bibitem [{\citenamefont {Liu}\ \emph {et~al.}(2020)\citenamefont {Liu} \emph {et~al.}}]{doi:10.1021}%
  \BibitemOpen
  \bibfield  {author} {\bibinfo {author} {\bibfnamefont {Y.}~\bibnamefont {Liu}} \emph {et~al.},\ }\bibfield  {title} {\bibinfo {title} {Coherent epitaxial semiconductor–ferromagnetic insulator {InAs/EuS} interfaces: {B}and alignment and magnetic structure},\ }\href {https://doi.org/10.1021/acsami.9b15034} {\bibfield  {journal} {\bibinfo  {journal} {ACS Appl. Mater. Interfaces}\ }\textbf {\bibinfo {volume} {12}},\ \bibinfo {pages} {8780} (\bibinfo {year} {2020})}\BibitemShut {NoStop}%
\bibitem [{\citenamefont {Ilse}\ \emph {et~al.}(2023)\citenamefont {Ilse}, \citenamefont {Sch\"utz},\ and\ \citenamefont {Goering}}]{PhysRevLett.131.036201}%
  \BibitemOpen
  \bibfield  {author} {\bibinfo {author} {\bibfnamefont {S.~E.}\ \bibnamefont {Ilse}}, \bibinfo {author} {\bibfnamefont {G.}~\bibnamefont {Sch\"utz}},\ and\ \bibinfo {author} {\bibfnamefont {E.}~\bibnamefont {Goering}},\ }\bibfield  {title} {\bibinfo {title} {Voltage {X}-ray reflectometry: A method to study electric-field-induced changes in interfacial electronic structures},\ }\href {https://doi.org/10.1103/PhysRevLett.131.036201} {\bibfield  {journal} {\bibinfo  {journal} {Phys. Rev. Lett.}\ }\textbf {\bibinfo {volume} {131}},\ \bibinfo {pages} {036201} (\bibinfo {year} {2023})}\BibitemShut {NoStop}%
\bibitem [{\citenamefont {Su}\ \emph {et~al.}(2012)\citenamefont {Su}, \citenamefont {Lee}, \citenamefont {Lin},\ and\ \citenamefont {Huang}}]{Roughness_comp}%
  \BibitemOpen
  \bibfield  {author} {\bibinfo {author} {\bibfnamefont {H.-C.}\ \bibnamefont {Su}}, \bibinfo {author} {\bibfnamefont {C.-H.}\ \bibnamefont {Lee}}, \bibinfo {author} {\bibfnamefont {M.-Z.}\ \bibnamefont {Lin}},\ and\ \bibinfo {author} {\bibfnamefont {T.-W.}\ \bibnamefont {Huang}},\ }\bibfield  {title} {\bibinfo {title} {A comparison between {X}-ray reflectivity and atomic force microscopy on the characterization of a surface roughness},\ }\href@noop {} {\bibfield  {journal} {\bibinfo  {journal} {Chin. J. Phys.}\ }\textbf {\bibinfo {volume} {50}},\ \bibinfo {pages} {291} (\bibinfo {year} {2012})}\BibitemShut {NoStop}%
\bibitem [{\citenamefont {Freitag}\ and\ \citenamefont {Clemens}(2001)}]{doi:10.1063/1.1332095}%
  \BibitemOpen
  \bibfield  {author} {\bibinfo {author} {\bibfnamefont {J.~M.}\ \bibnamefont {Freitag}}\ and\ \bibinfo {author} {\bibfnamefont {B.~M.}\ \bibnamefont {Clemens}},\ }\bibfield  {title} {\bibinfo {title} {Nonspecular {X}-ray reflectivity study of roughness scaling in {Si/Mo} multilayers},\ }\href {https://doi.org/10.1063/1.1332095} {\bibfield  {journal} {\bibinfo  {journal} {J. Appl. Phys.}\ }\textbf {\bibinfo {volume} {89}},\ \bibinfo {pages} {1101} (\bibinfo {year} {2001})}\BibitemShut {NoStop}%
\bibitem [{\citenamefont {Homma}\ \emph {et~al.}(2003)\citenamefont {Homma} \emph {et~al.}}]{SIMS_sputter}%
  \BibitemOpen
  \bibfield  {author} {\bibinfo {author} {\bibfnamefont {Y.}~\bibnamefont {Homma}} \emph {et~al.},\ }\bibfield  {title} {\bibinfo {title} {Evaluation of the sputtering rate variation in {SIMS} ultra-shallow depth profiling using multiple short-period delta layers},\ }\href {https://doi.org/https://doi.org/10.1002/sia.1568} {\bibfield  {journal} {\bibinfo  {journal} {Surf. Interface Anal.}\ }\textbf {\bibinfo {volume} {35}},\ \bibinfo {pages} {544} (\bibinfo {year} {2003})}\BibitemShut {NoStop}%
\bibitem [{\citenamefont {Hagmann}\ \emph {et~al.}(2020)\citenamefont {Hagmann} \emph {et~al.}}]{PhysRevB.101.245419}%
  \BibitemOpen
  \bibfield  {author} {\bibinfo {author} {\bibfnamefont {J.~A.}\ \bibnamefont {Hagmann}} \emph {et~al.},\ }\bibfield  {title} {\bibinfo {title} {Electron-electron interactions in low-dimensional {Si:P} delta layers},\ }\href {https://doi.org/10.1103/PhysRevB.101.245419} {\bibfield  {journal} {\bibinfo  {journal} {Phys. Rev. B}\ }\textbf {\bibinfo {volume} {101}},\ \bibinfo {pages} {245419} (\bibinfo {year} {2020})}\BibitemShut {NoStop}%
\bibitem [{\citenamefont {Hagmann}\ \emph {et~al.}(2018)\citenamefont {Hagmann} \emph {et~al.}}]{NIST_SiP}%
  \BibitemOpen
  \bibfield  {author} {\bibinfo {author} {\bibfnamefont {J.~A.}\ \bibnamefont {Hagmann}} \emph {et~al.},\ }\bibfield  {title} {\bibinfo {title} {{High resolution thickness measurements of ultrathin Si:P monolayers using weak localization}},\ }\href {https://doi.org/10.1063/1.4998712} {\bibfield  {journal} {\bibinfo  {journal} {Appl. Phys. Lett.}\ }\textbf {\bibinfo {volume} {112}},\ \bibinfo {pages} {043102} (\bibinfo {year} {2018})}\BibitemShut {NoStop}%
\bibitem [{\citenamefont {Keimer}\ \emph {et~al.}(2015)\citenamefont {Keimer}, \citenamefont {Kivelson}, \citenamefont {Norman}, \citenamefont {Uchida},\ and\ \citenamefont {Zaanen}}]{keimer2015quantum}%
  \BibitemOpen
  \bibfield  {author} {\bibinfo {author} {\bibfnamefont {B.}~\bibnamefont {Keimer}}, \bibinfo {author} {\bibfnamefont {S.~A.}\ \bibnamefont {Kivelson}}, \bibinfo {author} {\bibfnamefont {M.~R.}\ \bibnamefont {Norman}}, \bibinfo {author} {\bibfnamefont {S.}~\bibnamefont {Uchida}},\ and\ \bibinfo {author} {\bibfnamefont {J.}~\bibnamefont {Zaanen}},\ }\bibfield  {title} {\bibinfo {title} {From quantum matter to high-temperature superconductivity in copper oxides},\ }\href@noop {} {\bibfield  {journal} {\bibinfo  {journal} {Nature}\ }\textbf {\bibinfo {volume} {518}},\ \bibinfo {pages} {179} (\bibinfo {year} {2015})}\BibitemShut {NoStop}%
\bibitem [{\citenamefont {Logvenov}\ \emph {et~al.}(2009)\citenamefont {Logvenov}, \citenamefont {Gozar},\ and\ \citenamefont {Bozovic}}]{doi:10.1126/science.1178863}%
  \BibitemOpen
  \bibfield  {author} {\bibinfo {author} {\bibfnamefont {G.}~\bibnamefont {Logvenov}}, \bibinfo {author} {\bibfnamefont {A.}~\bibnamefont {Gozar}},\ and\ \bibinfo {author} {\bibfnamefont {I.}~\bibnamefont {Bozovic}},\ }\bibfield  {title} {\bibinfo {title} {High-temperature superconductivity in a single copper-oxygen plane},\ }\href {https://doi.org/10.1126/science.1178863} {\bibfield  {journal} {\bibinfo  {journal} {Science}\ }\textbf {\bibinfo {volume} {326}},\ \bibinfo {pages} {699} (\bibinfo {year} {2009})}\BibitemShut {NoStop}%
\bibitem [{\citenamefont {Li}\ \emph {et~al.}(2019)\citenamefont {Li} \emph {et~al.}}]{li2019superconductivity}%
  \BibitemOpen
  \bibfield  {author} {\bibinfo {author} {\bibfnamefont {D.}~\bibnamefont {Li}} \emph {et~al.},\ }\bibfield  {title} {\bibinfo {title} {Superconductivity in an infinite-layer nickelate},\ }\href@noop {} {\bibfield  {journal} {\bibinfo  {journal} {Nature}\ }\textbf {\bibinfo {volume} {572}},\ \bibinfo {pages} {624} (\bibinfo {year} {2019})}\BibitemShut {NoStop}%
\bibitem [{\citenamefont {Zeng}\ \emph {et~al.}(2022)\citenamefont {Zeng} \emph {et~al.}}]{doi:10.1126/sciadv.abl9927}%
  \BibitemOpen
  \bibfield  {author} {\bibinfo {author} {\bibfnamefont {S.}~\bibnamefont {Zeng}} \emph {et~al.},\ }\bibfield  {title} {\bibinfo {title} {Superconductivity in infinite-layer nickelate {La$_{1-x}$Ca$_x$NiO$_2$} thin films},\ }\href {https://doi.org/10.1126/sciadv.abl9927} {\bibfield  {journal} {\bibinfo  {journal} {Sci. Adv.}\ }\textbf {\bibinfo {volume} {8}},\ \bibinfo {pages} {eabl9927} (\bibinfo {year} {2022})}\BibitemShut {NoStop}%
\bibitem [{\citenamefont {Lee}\ \emph {et~al.}(2020)\citenamefont {Lee} \emph {et~al.}}]{10.1063/5.0005103}%
  \BibitemOpen
  \bibfield  {author} {\bibinfo {author} {\bibfnamefont {K.}~\bibnamefont {Lee}} \emph {et~al.},\ }\bibfield  {title} {\bibinfo {title} {{Aspects of the synthesis of thin film superconducting infinite-layer nickelates}},\ }\href {https://doi.org/10.1063/5.0005103} {\bibfield  {journal} {\bibinfo  {journal} {APL Mater.}\ }\textbf {\bibinfo {volume} {8}},\ \bibinfo {pages} {041107} (\bibinfo {year} {2020})}\BibitemShut {NoStop}%
\bibitem [{\citenamefont {Nomura}\ and\ \citenamefont {Arita}(2022)}]{Nomura_2022}%
  \BibitemOpen
  \bibfield  {author} {\bibinfo {author} {\bibfnamefont {Y.}~\bibnamefont {Nomura}}\ and\ \bibinfo {author} {\bibfnamefont {R.}~\bibnamefont {Arita}},\ }\bibfield  {title} {\bibinfo {title} {Superconductivity in infinite-layer nickelates},\ }\href {https://doi.org/10.1088/1361-6633/ac5a60} {\bibfield  {journal} {\bibinfo  {journal} {Rep. Prog. Phys.}\ }\textbf {\bibinfo {volume} {85}},\ \bibinfo {pages} {052501} (\bibinfo {year} {2022})}\BibitemShut {NoStop}%
\bibitem [{\citenamefont {Parzyck}\ \emph {et~al.}(2024)\citenamefont {Parzyck} \emph {et~al.}}]{parzyck2023absence}%
  \BibitemOpen
  \bibfield  {author} {\bibinfo {author} {\bibfnamefont {C.~T.}\ \bibnamefont {Parzyck}} \emph {et~al.},\ }\bibfield  {title} {\bibinfo {title} {Absence of $3a_0$ charge density wave order in the infinite-layer nickelate {NdNiO$_2$}},\ }\href {https://doi.org/10.1038/s41563-024-01797-0} {\bibfield  {journal} {\bibinfo  {journal} {Nat. Mater.}\ }\textbf {\bibinfo {volume} {23}},\ \bibinfo {pages} {486} (\bibinfo {year} {2024})}\BibitemShut {NoStop}%
\bibitem [{\citenamefont {Villar}\ \emph {et~al.}(2018)\citenamefont {Villar} \emph {et~al.}}]{ID16A}%
  \BibitemOpen
  \bibfield  {author} {\bibinfo {author} {\bibfnamefont {F.}~\bibnamefont {Villar}} \emph {et~al.},\ }\bibfield  {title} {\bibinfo {title} {Nanopositioning for the {ESRF} {ID16A} nano-imaging beamline},\ }\href {https://doi.org/10.1080/08940886.2018.1506234} {\bibfield  {journal} {\bibinfo  {journal} {Synchrotron Radiat. News}\ }\textbf {\bibinfo {volume} {31}},\ \bibinfo {pages} {9} (\bibinfo {year} {2018})}\BibitemShut {NoStop}%
\bibitem [{\citenamefont {Chantler}(2000)}]{Chantler2000}%
  \BibitemOpen
  \bibfield  {author} {\bibinfo {author} {\bibfnamefont {C.~T.}\ \bibnamefont {Chantler}},\ }\bibfield  {title} {\bibinfo {title} {{Detailed tabulation of atomic form factors, photoelectric absorption and scattering cross section, and mass attenuation coefficients in the vicinity of absorption edges in the soft X-Ray ($Z=$30–36, $Z=$60–89, $E=0.1$ keV – 10 keV), addressing convergence issues of earlier work}},\ }\href {https://doi.org/10.1063/1.1321055} {\bibfield  {journal} {\bibinfo  {journal} {J. Phys. Chem. Ref. Data}\ }\textbf {\bibinfo {volume} {29}},\ \bibinfo {pages} {597} (\bibinfo {year} {2000})}\BibitemShut {NoStop}%
\bibitem [{\citenamefont {Bj{\"{o}}rck}\ and\ \citenamefont {Andersson}(2007)}]{GenX}%
  \BibitemOpen
  \bibfield  {author} {\bibinfo {author} {\bibfnamefont {M.}~\bibnamefont {Bj{\"{o}}rck}}\ and\ \bibinfo {author} {\bibfnamefont {G.}~\bibnamefont {Andersson}},\ }\bibfield  {title} {\bibinfo {title} {{{\it GenX}: an extensible {X}-ray reflectivity refinement program utilizing differential evolution}},\ }\href {https://doi.org/10.1107/S0021889807045086} {\bibfield  {journal} {\bibinfo  {journal} {J. Appl. Crystallogr.}\ }\textbf {\bibinfo {volume} {40}},\ \bibinfo {pages} {1174} (\bibinfo {year} {2007})}\BibitemShut {NoStop}%
\bibitem [{\citenamefont {Goh}\ \emph {et~al.}(2005)\citenamefont {Goh}, \citenamefont {Oberbeck},\ and\ \citenamefont {Simmons}}]{Goh2005}%
  \BibitemOpen
  \bibfield  {author} {\bibinfo {author} {\bibfnamefont {K.~E.~J.}\ \bibnamefont {Goh}}, \bibinfo {author} {\bibfnamefont {L.}~\bibnamefont {Oberbeck}},\ and\ \bibinfo {author} {\bibfnamefont {M.~Y.}\ \bibnamefont {Simmons}},\ }\bibfield  {title} {\bibinfo {title} {Relevance of phosphorus incorporation and hydrogen removal for {Si:P} $\delta$-doped layers fabricated using phosphine},\ }\href {https://doi.org/https://doi.org/10.1002/pssa.200460764} {\bibfield  {journal} {\bibinfo  {journal} {Phys. Status Solidi A}\ }\textbf {\bibinfo {volume} {202}},\ \bibinfo {pages} {1002} (\bibinfo {year} {2005})}\BibitemShut {NoStop}%
\bibitem [{\citenamefont {Keizer}\ \emph {et~al.}(2015)\citenamefont {Keizer}, \citenamefont {Koelling}, \citenamefont {Koenraad},\ and\ \citenamefont {Simmons}}]{Keizer2015}%
  \BibitemOpen
  \bibfield  {author} {\bibinfo {author} {\bibfnamefont {J.~G.}\ \bibnamefont {Keizer}}, \bibinfo {author} {\bibfnamefont {S.}~\bibnamefont {Koelling}}, \bibinfo {author} {\bibfnamefont {P.~M.}\ \bibnamefont {Koenraad}},\ and\ \bibinfo {author} {\bibfnamefont {M.~Y.}\ \bibnamefont {Simmons}},\ }\bibfield  {title} {\bibinfo {title} {Suppressing segregation in highly phosphorus doped silicon monolayers},\ }\href {https://doi.org/10.1021/acsnano.5b06299} {\bibfield  {journal} {\bibinfo  {journal} {ACS Nano}\ }\textbf {\bibinfo {volume} {9}},\ \bibinfo {pages} {12537} (\bibinfo {year} {2015})}\BibitemShut {NoStop}%
\bibitem [{\citenamefont {Flechsig}\ \emph {et~al.}(2010)\citenamefont {Flechsig} \emph {et~al.}}]{doi:10.1063/1.3463200}%
  \BibitemOpen
  \bibfield  {author} {\bibinfo {author} {\bibfnamefont {U.}~\bibnamefont {Flechsig}} \emph {et~al.},\ }\bibfield  {title} {\bibinfo {title} {Performance measurements at the {SLS} {SIM} beamline},\ }\href {https://doi.org/10.1063/1.3463200} {\bibfield  {journal} {\bibinfo  {journal} {AIP Conf. Proc.}\ }\textbf {\bibinfo {volume} {1234}},\ \bibinfo {pages} {319} (\bibinfo {year} {2010})}\BibitemShut {NoStop}%
\bibitem [{\citenamefont {Staub}\ \emph {et~al.}(2008)\citenamefont {Staub} \emph {et~al.}}]{Staub:fh5385}%
  \BibitemOpen
  \bibfield  {author} {\bibinfo {author} {\bibfnamefont {U.}~\bibnamefont {Staub}} \emph {et~al.},\ }\bibfield  {title} {\bibinfo {title} {{Polarization analysis in soft {X}-ray diffraction to study magnetic and orbital ordering}},\ }\href {https://doi.org/10.1107/S0909049508019614} {\bibfield  {journal} {\bibinfo  {journal} {J. Synchrotron Radiat.}\ }\textbf {\bibinfo {volume} {15}},\ \bibinfo {pages} {469} (\bibinfo {year} {2008})}\BibitemShut {NoStop}%
\end{thebibliography}%
\end{document}